\documentclass[a4paper, 11pt]{article}
\usepackage{srcltx}
\usepackage{indentfirst}
\usepackage{graphicx}
\usepackage{overpic}
\usepackage{multirow}
\usepackage[dvips]{hyperref}
\usepackage[hang]{subfigure}
\usepackage{amsmath, amssymb, amsthm}
\usepackage{color}
\usepackage[mathscr]{euscript}

\newcommand\bbR{\mathbb{R}}
\newcommand\bbN{\mathbb{N}}
\newcommand\bxi{\boldsymbol{\xi}}
\newcommand\bx{\boldsymbol{x}}
\newcommand\bv{\boldsymbol{v}}
\newcommand\bu{\boldsymbol{u}}

\newcommand\bF{\boldsymbol{F}}

\newcommand\dd{\,\mathrm{d}}
\newcommand\He{\mathit{He}}
\newcommand\Kn{\mathit{Kn}}

\newcommand\pdd[1]{\dfrac{\partial}{\partial {#1}}}
\newcommand\pd[2]{\dfrac{\partial {#1}}{\partial {#2}}}
\newcommand\od[2]{\dfrac{\mathrm{d} {#1}}{\mathrm{d} {#2}}}

\numberwithin{equation}{section}
\graphicspath{{images/}}

{\theoremstyle{remark} \newtheorem{remark}{Remark}}

\title{Numerical Regularized Moment Method for High Mach Number Flow}

\author{Zhenning Cai\thanks{School of Mathematical Sciences, Peking
    University, Beijing, China, email: {\tt harecat@gmail.com}.},~~ Ruo
  Li\thanks{CAPT, LMAM \& School of Mathematical Sciences, Peking
    University, Beijing, China, email: {\tt rli@math.pku.edu.cn}.}~~ and
  Yanli Wang\thanks{School of Mathematical Sciences, Peking
    University, Beijing, China, email: {\tt wangyanliwyl@gmail.com}.}}

\begin{document}
\maketitle
% vim: tw=70:spell
\begin{abstract}
  This paper is a continuation of our earlier work \cite{NRxx} in
  which a numerical moment method with arbitrary order of moments was
  presented. However, the computation may break down during the
  calculation of the structure of a shock wave with Mach number $M_0
  \geqslant 3$. In this paper, we concentrate on the regularization of
  the moment systems. First, we apply the Maxwell iteration to the
  infinite moment system and determine the magnitude of each moment
  with respect to the Knudsen number. After that, we obtain the
  approximation of high order moments and close the moment systems by
  dropping some high-order terms. Linearization is then performed to
  obtain a very simple regularization term, thus it is very convenient
  for numerical implementation. To validate the new regularization,
  the shock structures of low order systems are computed with different
  shock Mach numbers.

\vspace*{4mm}
\noindent {\bf Keywords:} Boltzmann-BGK equation; Maxwellian iteration;
 Regularized moment equations
\end{abstract}

\section{Introduction} \label{sec:intro}
In the field such as high altitude flight and microscopic flows, gas
is considered to be very rarefied and outside the hydrodynamic regime.
In this case, usual fluid models such as Euler equations and
Navier-Stokes-Fourier system will fail when the rarefied effect is
significant. The moment method, which was first proposed by Grad
\cite{Grad}, is focused on the description of the rarefied gases using
a small number of variables. Almost all moment methods are derived
from the Boltzmann equation which is regarded to be able to capture
the rarefied effects accurately. In \cite{NRxx}, a special expansion
of the distribution functions is adopted to make it possible the solve
the associated Grad-type moment equations numerically without the
explicit expressions of the system. Then, we followed \cite
{Struchtrup2003} and numerically regularized the system using the
technique of a modified Chapman-Enskog expansion. In \cite{NRxx}, it
has been verified numerically that a smooth shock structure with Mach
number $M_0 = 2$ can be obtained by solving the R20 equations with a
Riemann problem until the steady state. However, it was found in our
numerical experiments that if we set the shock Mach number $M_0
\geqslant 3$, the computation will break down with negative
temperature appearing inside the shock wave before a steady structure
of the shock wave is formed, which is possibly caused by the
non-hyperbolic nature of the moment system.

In this paper, we present a new regularization method which is able to
produce smooth profile for large Mach number shock waves and low order
moment systems. The idea originates from the order-of-magnitude method
\cite{Struchtrup,Struchtrup2004}, where the order of magnitude for
each moment with respect to the Knudsen number is investigated in
order to obtain a transport system with a specified order of accuracy.
Additionally, from the computational perspective, a conservative form
of moment equations is preferred, so we put this idea into the
framework of \cite{NRxx} and derive a uniform expression of the
regularization terms for all moment systems.

As the first step, we derive the analytical form of the moment equations
using the same set of moments as in \cite{NRxx}. Once the moment
equations are given explicitly, we find that only conservative
variables and the moments within four successive orders appear in each
equation. With the help of Maxwellian iteration \cite{Ikenberry}, the
order of magnitude can be obtained for each moment, and this skill has
been used in \cite{Reitebuch}. The closure of the moment system is
achieved using a similar skill as in \cite{Reitebuch, Struchtrup2005}.
We approximate the $(M+1)$-st order moments by removing all terms with
higher orders of magnitude than the leading order in the corresponding
equation to get the closed system of all moments with orders lower
than $M$.  Eventually, a parabolic system is explicitly obtained.

The resulting regularization term is somewhat complicated and is
simplified using the technique of linearization for the sake of
convenient numerical implementation. As in \cite{Struchtrup2003}, the
fluid is considered to be in the vicinity of velocity-free
equilibrium states, thus the derivatives are small. Dropping the
terms which are nonlinear in small values, the remaining linear part
turns to be very compact. In the 1D case, it is obvious that the
regularization introduces additional diffusion to the $M$-th order
moments. With the simplified regularization term, the numerical
investigation of the shock tube problem shows the convergence to the
Boltzmann-BGK equation in moments. And it is numerically demonstrated
that smooth shock profiles can be obtained for large Mach numbers.

The layout of this paper is as follows: in Section 2, an overview of
Boltzmann-BGK model and our discretization of the distribution
function is introduced as some preliminaries. In section 3, the
details of the new regularization method are presented. In section 4,
we present two numerical examples to make comparisons between results
for different moment equations, different Knudsen numbers and
different Mach numbers. At last, some concluding remarks will be given
in Section 5.

%%% Local Variables: 
%%% mode: latex
%%% TeX-master: "article"
%%% End: 

% vim: tw=70:spell
\section{Preliminaries}
\subsection{The Boltzmann-BGK model}
In the mesoscopic view, the gas can be characterized by the
distribution function $f(t, \bx, \bxi)$, where $t$, $\bx$ and $\bxi$
stand for the time, the spatial position and the particle velocity
respectively, and $\bx, \bxi \in \bbR^D$, $D \leqslant 3$. The
macroscopic quantities including the density $\rho$, the velocity
$\bu$ and the temperature $T$ can be related with $f$ by
\begin{equation} \label{eq:moments}
\begin{split} \rho(t, \bx) &= \int_{\bbR^D} f(t, \bx, \bxi) \dd \bxi, \\
\rho(t, \bx) \bu(t, \bx) &=
  \int_{\bbR^D} \bxi f(t, \bx, \bxi) \dd \bxi, \\
\rho(t, \bx) |\bu(t, \bx)|^2 + D \rho(t, \bx) R T(t, \bx)
  &= \int_{\bbR^D} |\bxi|^2 f(t, \bx, \bxi) \dd \bxi,
\end{split}
\end{equation}
where $R$ is the gas constant. As usual, we use $\theta(t, \bx) = RT(t,
\bx)$ to simplify the notation. Since $R$ is a constant, $\theta$ can
also be considered as the temperature in the non-dimensional case.

The Boltzmann-BGK model is a simplification of the classical Boltzmann
equation, which uses a relaxation term instead of the binary collision
operator. The Boltzmann-BGK equation reads
\begin{equation} \label{eq:BGK}
  \pd{f}{t} + \bxi \cdot \nabla_{\bx} f = \frac{1}{\tau} (f_M - f),
\end{equation}
where $\tau$ is the relaxation time, and $f_M$ is the local Maxwellian
defined as
\begin{equation}
f_M(t, \bx, \bxi) = \frac{\rho(t, \bx)}
  {[2\pi \theta(t, \bx)]^{D/2}}
  \exp \left(
    -\frac{|\bxi - \bu(t, \bx)|^2}{2\theta(t, \bx)}
  \right).
\end{equation}
By multiplying the Boltzmann equation by $(1,\bxi, |\bxi|^2/2)^T$,
and integrating both sides over $\bbR^D$ with respect to $\bxi$, we
get the conservation laws as
\begin{align}
\label{eq:drho_dt}
  &\od {\rho}{t} + \rho \sum_{k=1}^D \pd{u_k}{x_k} = 0, \\
\label{eq:du_dt}
  &\rho \od{ u_i}{ t} + \pd{p}{x_i}
    + \sum_{k=1}^D \pd{\sigma_{ik}}{x_k} = 0, \\
\label{eq:dtheta_dt}
  &\frac{D}{2}\rho \od{\theta}{ t} + \sum_{k=1}^D \pd{q_k}{x_k}
    = -\sum_{i=1}^D \sum_{j=1}^D p_{ij} \pd{u_i}{x_j},
\end{align}
where $\frac{\mathrm{d}}{\mathrm{d}t}$ is the  material derivative,
$p_{ij}$, $\sigma_{ij}$ and $q_k$ are the pressure tensor, the stress
tensor and the heat flux, respectively. The precise definitions are
\begin{equation} \label{eq:P}
\begin{split}
  \od{\psi}{t}& = \pd{\psi}{t} + \bu \cdot \nabla_{\bx} \psi,
    \quad \psi = \rho, u_i, \theta,\\
  p_{ij} &= \int_{\bbR^D} (\xi_i-u_i)(\xi_j-u_j) f \dd \bxi,
    \quad \sigma_{ij} = p_{ij} - p \delta_{ij}, \quad p = \rho \theta, \\
  q_k &= \frac{1}{2} \int_{\bbR^D}|\bxi-\bu|^2(\xi_k-u_k) f \dd \bxi,
    \quad i,j,k = 1,\cdots,D, \\
\end{split}
\end{equation} 
where we have used the ideal gas law in $p = \rho \theta$.

\subsection{Discretization of the distribution function}
\label{sec:discretization}
Suppose $\bx$ and $t$ are fixed, we expand the distribution function
into Hermite functions as in \cite{Grad}
\begin{equation} \label{eq:expansion}
f(\bxi) = \sum_{\alpha \in {\bbN^D}} f_{\alpha}
  \mathcal{H}_{\theta,\alpha}(\bv),
\end{equation}
where $\alpha = (\alpha_1, \cdots, \alpha_D)$ is a $D$-dimensional
multi-index, and
\begin{equation} \label{eq:v}
\bv = \frac{\bxi - \bu}{\sqrt{\theta}}.
\end{equation}
The basis functions $\mathcal{H}_{\theta,\alpha}$ are chosen as
\begin{equation} \label{eq:H}
\mathcal{H}_{\theta,\alpha}(\bv) = \prod_{d=1}^D
  \frac{1}{\sqrt{2\pi}} \theta^{-\frac{\alpha_d + 1}{2}}
  \He_{\alpha_d}(v_d) \exp \left(-\frac{v_d^2}{2}\right),
\end{equation}
where $\He_{\alpha_d}$ is the Hermite polynomial defined by
\begin{equation} \label{eq:He}
\He_n(x) = (-1)^n \exp\left(\frac{x^2}{2}\right)
  \frac{\mathrm{d}^n}{\mathrm{d}x^n} \exp\left(-\frac{x^2}{2}\right).
\end{equation}
$\He_n$ is assumed to be zero if $n$ is negative. Thus $\mathcal{H}_
{\theta,\alpha}$ is zero if any component of $\alpha$ is negative.
Some useful properties of the Hermite polynomials are listed in
Appendix \ref{sec:Hermite}. It has been derived in \cite{NRxx}
from these properties that the following relations hold
\begin{equation} \label{eq:low_order}
f_0 = \rho, \quad f_{e_i} = 0, \quad \sum_{d=1}^D f_{2e_d} = 0,
  \qquad i = 1,\cdots,D.
\end{equation}
The stress tensor and heat flux can also be expressed in a simple
form:
\begin{equation}
\begin{gathered}
\sigma_{ij} = f_{e_i + e_j}, \quad \sigma_{jj} = 2f_{2e_j}, \qquad
  i,j = 1,\cdots,D, \quad i \neq j, \\
q_k = 2f_{3e_k} + \sum_{d=1}^D f_{2e_d + e_k}, \qquad k = 1,\cdots,D.
\end{gathered}
\end{equation}

In fact, \eqref{eq:expansion} defines a set of moments $\mathcal{M} =
\{f_{\alpha}\}_{\alpha \in \bbN^D}$, which will result in an
``infinite moment system'' if we put \eqref{eq:expansion} into
\eqref{eq:BGK}. In order to get a system with finite number of equations, we
choose a positive integer $M \geqslant 3$ and consider only a subset
of $\mathcal{M}$ which contains $f_{\alpha}$ with $|\alpha| \leqslant
M$. If we simply force the remaining moments to be zero, the Grad-type
system associated with the moment set $\{f_{\alpha}\}_{|\alpha|
\leqslant M}$ is obtained.

In \cite{NRxx}, the same set of moments has been used and we have
already constructed a numerical algorithm for solving the associated
Grad-type systems. In the remaining part of this paper, we will focus
on the regularization of such moment systems.

\section{Regularization of the moment method}
Using a similar technique as in \cite{Struchtrup2003}, one possible
regularization for the Grad-type systems with moments
$\{f_{\alpha}\}_{|\alpha| \leqslant M}$ has been introduced in \cite
{NRxx}, where a numerical regularization algorithm is proposed without
deriving the analytical expressions of the regularization terms.
However, such regularization can cause breakdown of the computation
due to the appearance of negative temperature while solving the shock
structure with Mach number $M_0 \geqslant 3$. This is possibly caused
by the non-hyperbolic nature of the moment system. In this section, a
new regularization method is proposed following the idea of Struchtrup
\cite{Struchtrup2004}.

\subsection{Maxwellian iteration} \label{sec:iter_result}
The Maxwellian iteration was introduced by Ikenberry and Truesdell in
\cite{Ikenberry} as a technique to derive NSF and Burnett equations
from the moment equations. Later in \cite{Reitebuch}, it is used as a
tool to analyse the order of magnitude of each moment and to derive
equations for extended thermodynamics, which is known as the COET
method. Below we apply Maxwellian iteration to the moment set
$\mathcal{M}$, and give a uniform description on the orders of
magnitude for the moments.

In order to perform the Maxwellian iteration, we first need to derive
the explicit expressions of the infinite moment equations. This can be
done by substituting the distribution function $f$ in \eqref{eq:BGK}
with its expansion \eqref{eq:expansion}. After some calculation, both
sides of \eqref{eq:BGK} can be expanded into Hermite series. Then, we
match the coefficient of each basis function and then the moment
equations can be obtained. The details can be found in Appendix
\ref{sec:moment_eqs}. Here, we write the moment equations \eqref
{eq:moment_eqs} as the following form with only one moment on the left
of each equality:
\begin{equation} \label{eq:iter}
\begin{split}
  f_{\alpha} = &-\tau \bigg\{
    \pd{f_{\alpha}}{t} + \sum_{d=1}^D \pd{u_d}{t} f_{\alpha-e_{d}}
    + \frac{1}{2} \pd{\theta}{t} \sum_{d =1}^D f_{\alpha-2e_d} \\
  & \qquad +\sum_{j=1}^D \bigg[ \left(
    \theta \pd{f_{\alpha-e_j}}{x_j} + u_j \pd{f_{\alpha}}{x_j}
    + (\alpha_j+1) \pd{f_{\alpha+e_j}}{x_j}
  \right) \\
  & \qquad +\sum_{d=1}^D \pd{u_d}{x_j} \left(
    \theta f_{\alpha-e_d-e_j} + u_j f_{\alpha - e_d}
    + (\alpha_j+1) f_{\alpha-e_d+e_j}
  \right) \\
  & \qquad +\frac{1}{2} \pd{\theta}{x_j} \sum_{d=1}^D \left(
    \theta f_{\alpha-2e_d-e_j} + u_j f_{\alpha-2e_d} +
    (\alpha_j+1) f_{\alpha-2e_d+e_j}
  \right) \bigg] \bigg\}, \quad |\alpha| \geqslant 2.
\end{split}
\end{equation}
Note that the cases of $|\alpha| = 0$ and $|\alpha| = 1$ are not
included, since when $|\alpha| = 0$, the collision term is zero and
the form with $f_0$ on the left hand side does not exist, and
$f_{\alpha} \equiv 0$ when $|\alpha| = 1$ following \eqref
{eq:low_order}.

The equation \eqref{eq:iter} can be taken as an iterative scheme
\begin{equation} \label{eq:iter_scheme}
f_{\alpha}^{(n+1)} = -\tau \mathcal{G}_{\alpha} \left(
  f_{\beta}^{(n)} \mid \beta\in\bbN^D
\right),
  \quad \forall \alpha \in \bbN^D \text{ and } |\alpha| \geqslant 2,
  \qquad n = 0, 1, 2, \cdots
\end{equation}
with the initial value to be the Maxwellian
\begin{equation} \label{eq:initial}
f_0^{(0)} = \rho, \quad f_{\alpha}^{(0)} = 0,
  \quad \forall |\alpha| \geqslant 1.
\end{equation}
During the iteration, $f_0$ and $f_{e_j}$ are never changed since they
never appear on the left hand side of \eqref{eq:iter}. $\bu$ and
$\theta$ do not change with $n$ either. Thus, the operator
$\mathcal{G}_{\alpha}$ in \eqref{eq:iter_scheme} can be considered as
a linear operator according to its analytical form \eqref{eq:iter}.
For a simpler notation, we define the following vectors:
\begin{equation}
\bF_M^{(n)} = (f_{\alpha}^{(n)})_{|\alpha| = M}, \quad
\bF^{(n)} = (\bF_0^{(n)}, \bF_1^{(n)}, \bF_2^{(n)}, \cdots).
\end{equation}
Here $\bF^{(n)}$ is an infinite dimensional vector, but we will reveal
that only finite number of its components are nonzero. Thus, \eqref
{eq:iter_scheme} can be written as
\begin{equation}
\bF^{(n+1)} = -\tau \mathcal{G}(\bF^{(n)}),
\end{equation}
and precisely, it is a system as
\begin{equation} \label{eq:iter_alpha}
f_{\alpha}^{(n+1)} = -\tau \mathcal{G}_{\alpha} \left(
  \bF_{|\alpha|-3}^{(n)}, \bF_{|\alpha|-2}^{(n)},
  \bF_{|\alpha|-1}^{(n)}, \bF_{|\alpha|}^{(n)},
  \bF_{|\alpha|+1}^{(n)}
\right), \quad |\alpha| \geqslant 2.
\end{equation}

Now we start the iteration, and the first two steps will be concretely
presented as below.

\paragraph{The first step of iteration} As the first step, we start
from the initial values and put \eqref{eq:initial} into
\eqref{eq:iter_scheme}. Noting that most terms in the right hand side
of \eqref{eq:iter_scheme} are zero, we have
\begin{equation} \label{eq:first_iter}
\begin{split}
& f_{2e_j}^{(1)} = -\tau \left(
  \frac{1}{2} \rho \pd{\theta}{t} + \rho \theta \pd{u_j}{x_j}
    + \frac{1}{2} \rho \bu \cdot \nabla_{\bx} \theta
\right), \\
& f_{e_i + e_j}^{(1)} = -\tau \rho \theta
  \left( \pd{u_i}{x_j} + \pd{u_j}{x_i} \right), \quad i \neq j, \\
& f_{2e_i + e_j}^{(1)} =
  -\frac{1}{2} \tau \rho \theta \pd{\theta}{x_j}, \\
& f_{e_i + e_j + e_k}^{(1)} = 0, \quad i \neq j \neq k, \\
& f_{\alpha}^{(1)} = 0, \quad |\alpha| \geqslant 4,
\end{split}
\end{equation}
where all $f_0$'s are replaced by the density $\rho$. Taking $\tau$ as
a small quantity, all moments produced in the first iteration are not
larger than $O(\tau)$. Thus we have
\begin{equation}
\bF^{(1)} = \bF^{(0)} - \tau \widetilde{\bF}^{(1)},
  \quad \widetilde{\bF}^{(1)} = O(1).
\end{equation}

\paragraph{The second step of iteration} Due to the excessive
complexity of the expressions, the detailed formulas in the second
step of the iteration are not presented while the orders of magnitude
can be observed. Since $\mathcal{G}$ is a linear operator, we have
\begin{equation}
\bF^{(2)} = -\tau \mathcal{G}(\bF^{(1)})
  = -\tau \mathcal{G}(\bF^{(0)})
    +\tau^2 \mathcal{G}(\widetilde{\bF}^{(1)})
  = \bF^{(1)} + \tau^2 \mathcal{G}(\widetilde{\bF}^{(1)}).
\end{equation}
Thus only second order terms are added in this step of iteration. Due
to \eqref{eq:first_iter}, the moments $f_{\alpha}^{(2)}$ with
$|\alpha| = 2, 3$ are not larger than $O(\tau)$, and the moments with
$|\alpha| \geqslant 4$ are no larger than $O(\tau^2)$. Furthermore, the
moments with $|\alpha| \geqslant 7$ are zeros, which can be revealed by
\eqref{eq:iter_alpha} and the last equation in \eqref{eq:first_iter}.

Let us go one step closer. For $|\alpha| = 3$, since
\begin{equation} \label{eq:alpha=3}
f_{\alpha}^{(2)} = -\tau \sum_{j=1}^D \theta
  \pd{f_{\alpha-e_j}^{(1)}}{x_j} + \cdots, \quad |\alpha| = 3,
\end{equation}
$f_{\alpha}^{(2)}$ with $|\alpha| = 3$ is known to be no less than
$O(\tau^2)$. More precisely, we have
\begin{equation} \label{eq:f3}
f_{3e_k}^{(2)} = O(\tau), \quad
f_{2e_i + e_j}^{(2)} = O(\tau), \quad
f_{e_i + e_j + e_k}^{(2)} = O(\tau^2), \qquad i \neq j \neq k.
\end{equation}
For $|\alpha| = 4$, we have
\begin{equation} \label{eq:alpha=4}
f_{\alpha}^{(2)} = -\tau \sum_{j=1}^D \sum_{d=1}^D
  \pd{u_d}{x_j} \theta f_{\alpha-e_d-e_j}^{(1)} + \cdots
  = O(\tau^2), \quad |\alpha| = 4.
\end{equation}
For $|\alpha| = 5, 6$, since $D \leqslant 3$, $f_{\alpha}^{(2)}$ can
be estimated as
\begin{equation} \label{eq:alpha=56}
f_{\alpha}^{(2)} = -\tau \sum_{j=1}^D \frac{1}{2} \pd{\theta}{x_j}
  \sum_{d=1}^D \theta f_{\alpha-2e_d-e_j}^{(1)} + \cdots
  = O(\tau^2), \quad |\alpha| = 5, 6.
\end{equation}

\paragraph{The final conclusion} The technique used here can be
applied to further iterations recursively. It is then found by
induction that for any positive integer $n$, one has
\begin{equation} \label{eq:iter_F}
\begin{gathered}
\bF^{(n)} = \bF^{(n-1)} + \tau^n \mathcal{G}(\widetilde{\bF}^{(n)}),
  \quad \widetilde{\bF}^{(n)} = O(1).
\end{gathered}
\end{equation}
And for $|\alpha| \geqslant 1 + 3n$, $f_{\alpha}^{(n)}$ is zero. This
implies that for any $\alpha$ and $n$, $f_{\alpha}^{(n)}$ is never
larger than $O(\tau^{\lceil |\alpha| / 3 \rceil})$. Moreover, a
careful investigation similar as \eqref {eq:alpha=4} and
\eqref{eq:alpha=56} gives
\begin{equation} \label{eq:magnitude}
f_{\alpha}^{(n)} = O(\tau^{\lceil |\alpha| / 3 \rceil}),
  \qquad \forall \alpha \in \bbN^D \text{ and } |\alpha| \geqslant 4,
  \quad n \geqslant |\alpha| / 3.
\end{equation}
For $|\alpha| \leqslant 3$, detailed results have been given in \eqref
{eq:low_order}, \eqref{eq:first_iter} and \eqref{eq:f3}.

\begin{remark}
The equation \eqref{eq:iter_F} indicates that for any moment
$f_{\alpha}$, the leading order term is never changed once it is
obtained. Thus the leading order terms of the stress tensor
$\sigma_{ij}$ and the heat flux $q_k$ can be derived by
\eqref{eq:first_iter} as
\begin{gather}
\label{eq:Fourier}
q_k = 2f_{3e_k} + \sum_{d=1}^D f_{e_k + 2e_d}
  = -\frac{D + 2}{2} \tau \rho \theta \pd{\theta}{x_k} + O(\tau^2),
  \qquad k = 1, \cdots, D, \\
\label{eq:sigma_ij}
\sigma_{ij} = f_{e_i + e_j} = -\tau \rho \theta
  \left( \pd{u_i}{x_j} + \pd{u_j}{x_i} \right) + O(\tau^2), \qquad
  i,j = 1,\cdots,D, \quad i \neq j, \\
\label{eq:sigma_jj}
\sigma_{jj} = 2f_{2e_j} = -\tau \left[
  \rho \left( \pd{\theta}{t} + \bu \cdot \nabla_{\bx} \theta \right)
  + 2\rho \theta \pd{u_j}{x_j} \right] + O(\tau^2),
  \qquad j = 1, \cdots, D.
\end{gather}
As expected, the Fourier law is deduced as \eqref{eq:Fourier}. Using
\eqref{eq:dtheta_dt}, \eqref{eq:sigma_jj} can be further simplified as
\begin{equation} \label{eq:NS}
\begin{split}
\sigma_{jj} & = -2\tau \left[
  -\frac{1}{D} \left(
    \sum_{i=1}^D \sum_{k=1}^D (\rho \theta \delta_{ik} + \sigma_{ik})
      \pd{u_i}{x_k} + \sum_{k=1}^D \pd{q_k}{x_k}
  \right) + \rho \theta \pd{u_j}{x_j}
\right] + O(\tau^2) \\
& = -2 \tau \rho \theta \left(
  \pd{u_j}{x_j} - \frac{1}{D} \sum_{i=1}^D \pd{u_i}{x_i}
\right) + O(\tau^2),
\end{split}
\end{equation}
where we have used the fact that $\sigma_{ik} = O(\tau)$, $q_k =
O(\tau)$. The equations \eqref{eq:sigma_ij} and \eqref{eq:NS} can be
written uniformly as
\begin{equation} \label{eq:Navier-Stokes}
\sigma_{ij} = -2\tau \rho \theta \pd{u_{\langle i}}{x_{j \rangle}}
  + O(\tau^2), \quad i,j = 1,\cdots,D,
\end{equation}
which is the Navier-Stokes law. It is obvious that the above equations
and \eqref{eq:Fourier} yield a Prandtl number $\mathrm{Pr} = 1$,
which agrees with the common knowledge that the BGK equation produces
the incorrect Prandtl number $1$. This is a validation of the
Maxwellian iteration.

Actually, the Maxwellian iteration can be considered equivalent to the
Chapman-Enskog expansion, which also gives successive order of the
distribution function (see e.g. \cite{Struchtrup}). Here the
Maxwellian iteration is much easier to use than the Chapman-Enskog
expansion, though the latter one is able to give the same results.
\end{remark}

\begin{remark}
  It is possible in our calculation that some lower order terms will
  cancel each other and only high order terms remain in the iteration.
  That means some $f_{\alpha}$ with $\alpha \geqslant 4$ may have a
  smaller order of magnitude $o(\tau^{\lceil |\alpha| / 3 \rceil})$.
  However, the analysis has depicted the essential trend of the
  variation of the moments' magnitudes, which can be validated by
  comparing \eqref{eq:magnitude} with the tables in \cite{Reitebuch}.
  Meanwhile, the conciseness of \eqref{eq:magnitude} is very helpful
  to our later use.
\end{remark}

\subsection{The moment closure} \label{sec:iteration}
In order to close the moment system, as stated in Section
\ref{sec:discretization}, we first choose a positive integer $M
\geqslant 3$ and discard all equations containing the term $\partial
f_{\alpha} / \partial t$ with $|\alpha| > M$. Then, since $f_{\alpha}$
with $|\alpha| = M + 1$ remains in the system, it is to be substituted
by some expression consists of lower order moments only.

This can be done by using \eqref{eq:iter} and removing the high order
terms. With the help of \eqref{eq:magnitude}, the equation \eqref
{eq:iter} can be reformulated as 
\begin{equation} \label{eq:iter_s}
\begin{split}
f_{\alpha} =  &-\tau \Bigg\{ \left(
  \sum_{d=1}^D \pd{u_d}{t} f_{\alpha-e_d} +
  \frac{1}{2} \pd{\theta}{t} \sum_{d=1}^D f_{\alpha-2e_d}
\right) \\
& \qquad +\sum_{j=1}^D \Bigg[ \left(
  \theta \pd{f_{\alpha-e_j}}{x_j} +
  \sum_{d=1}^D \pd{u_d}{x_j}
    (\theta f_{\alpha-e_d-e_j} + u_j f_{\alpha-e_d})
\right) \\
& \qquad {} + \frac{1}{2} \pd{\theta}{x_j}
  \sum_{d=1}^D \left(
    \theta f_{\alpha-2e_d-e_j} + u_j f_{\alpha-2e_d}
    + (\alpha_j+1)f_{\alpha-2e_d +e_j}
  \right)
\Bigg] \Bigg\} + h.o.t.,
\end{split}
\end{equation}
where ``$h.o.t.$'' stands for high order terms, and it will not be
explicitly written later on. Note that
\begin{equation}
\od{u_d}{t} = \pd{u_d}{t} + \sum_{j=1}^D u_j \pd{u_d}{x_j}, \quad
\od{\theta}{t} = \pd{\theta}{t} + \sum_{j=1}^D u_j\pd{\theta}{x_j}.
\end{equation}
Putting them into \eqref{eq:iter_s}, we get
\begin{equation} \label{eq:f_alpha}
\begin{split}
f_{\alpha} =  &-\tau \Bigg\{ \left(
  \sum_{d=1}^D \od{u_d}{t} f_{\alpha-e_d} +
  \frac{1}{2} \od{\theta}{t} \sum_{d =1}^D f_{\alpha-2e_d}
\right) + \sum_{j=1}^D \Bigg[
  \sum_{d=1}^D \pd{u_{d}}{x_{j}} \theta f_{\alpha-e_d-e_j} \\
& \qquad {} + \theta \pd{f_{\alpha-e_j}}{x_j} +
  \frac{1}{2} \pd{\theta}{x_j} \sum_{d=1}^D
    (\theta f_{\alpha-2e_d-e_j} + (\alpha_j+1)f_{\alpha-2e_d +e_j})
\Bigg] \Bigg\}. \\
\end{split}
\end{equation}
Substituting the material derivatives by the equations \eqref{eq:du_dt}
and \eqref{eq:dtheta_dt}, \eqref{eq:f_alpha} is reformulated as
\begin{equation} \label{eq:f_alpha1}
\begin{split}
f_{\alpha} & = \tau \Bigg\{
  \frac{1}{\rho} \sum_{d=1}^D \sum_{j=1}^D \pd{p_{dj}}{x_j} f_{\alpha-e_d} +
  \frac{1}{D\rho} \left[
    \sum_{j=1}^D \left(
      \pd{q_j}{x_j} + \sum_{d=1}^D p_{dj} \pd{u_d}{x_j}
    \right)
  \right] \sum_{d=1}^D f_{\alpha-2e_d} - \\
& \quad \sum_{j=1}^D \left[
  \theta \pd{f_{\alpha-e_j}}{x_j} + \sum_{d=1}^D \left(
    \pd{u_d}{x_j} \theta f_{\alpha-e_d-e_j}
    + \frac{1}{2} \pd{\theta}{x_j}
      (\theta f_{\alpha-2e_d-e_j} + (\alpha_j+1) f_{\alpha-2e_d +e_j})
  \right)
\right] \Bigg\}. \\
\end{split}
\end{equation}
Recall that $\sigma_{ij}$ and $q_j$ have the order of magnitude
$O(\tau)$, and $p_{ij} = p \delta_{ij} + \sigma_{ij}$. The parts
containing $\sigma_{ij}$ and $q_j$ can also be discarded from \eqref
{eq:f_alpha1}. After that, one obtains
\begin{equation}
\begin{split}
f_{\alpha} &= \tau \Bigg\{
  \frac{1}{\rho} \sum_{j=1}^D \pd{p}{x_j} f_{\alpha-e_j}
  + \frac{\theta}{D} \left( \sum_{j=1}^D \pd{u_j}{x_j} \right)
    \sum_{d=1}^D f_{\alpha-2e_d}
  - \sum_{j=1}^D \Bigg[ \theta \pd{f_{\alpha-e_j}}{x_j} \\
& \qquad {} + \sum_{d=1}^D \left(
  \pd{u_d}{x_j} \theta f_{\alpha-e_d-e_j}
  + \frac{1}{2} \pd{\theta}{x_j}
    (\theta f_{\alpha-2e_d-e_j} + (\alpha_j+1) f_{\alpha-2e_d+e_j})
  \right)
\Bigg] \Bigg\}.
\end{split}
\end{equation}
This equation can be further simplified by coupling it with \eqref
{eq:Fourier} and \eqref{eq:Navier-Stokes}, and then dropping small
terms. The final result is
\begin{equation} \label{eq:reg_term}
\begin{split}
f_{\alpha} &= \tau \left(
  \frac{1}{\rho} \sum_{j=1}^D \pd{p}{x_j} f_{\alpha-e_j}
  - \sum_{j=1}^D \theta \pd{f_{\alpha-e_j}}{x_j}
\right) \\
& \qquad {} + \frac{1}{\rho} \sum_{j=1}^D \sum_{d=1}^D \left[
  \frac{1}{2} \sigma_{ij} f_{\alpha-e_d-e_j} +
  \frac{1}{(D+2)\theta} q_j (\theta f_{\alpha-2e_d-e_j}
    + (\alpha_j+1) f_{\alpha-2e_d+e_j})
\right].
\end{split}
\end{equation}
The equation \eqref{eq:reg_term} will be used for all $|\alpha| = M +
1$, and we ultimately obtain a closed parabolic system for the moment
set $\{f_{\alpha}\}_{|\alpha| \leqslant M}$.

\begin{remark}
As is well known, the most serious deficiency of the BGK collision
operator is that it predicts an incorrect Prandtl number. Therefore,
some other models such as the ES-BGK \cite{Holway} and Shakhov \cite
{Shakhov} models are proposed as a remedy. Until now, these models are
known to be very accurate in most cases. All these models have a
unified form of the collision term:
\begin{equation}
Q(f) = \bar{\nu} (G - f),
\end{equation}
where $\bar{\nu}$ is the average collision frequency, and $G$ is some
pseudo-equilibrium. The discretization of such collision operator has
been discussed in \cite{NRxx}. Here we emphasize that models with such
form can always be easily written as an iteration scheme like \eqref
{eq:iter_s} due to the existence of the term $-f$ in the collision
operator. Thus we can still use Maxwellian iteration to analyse the
order of magnitude for each moment.
\end{remark}

\subsection{Linearization of the regularization terms}
Once the regularization term \eqref{eq:reg_term} is constructed, the
system is closed. However, recalling that such moment systems are
mainly used for computation, it is clear that \eqref{eq:reg_term} is
not concise enough for implementation in numerical
simulation. Therefore, we are going to linearize \eqref{eq:reg_term}
and make its expression neater. The similar way has been used in
\cite{Torrilhon2006} for simplified numerical schemes.

The linearization will be taken in the neighbourhood of a
velocity-free Maxwellian. Suppose the radius $\epsilon$ of the
neighbourhood is small and
\begin{equation} \label{eq:linearize}
\begin{split}
& \rho = \rho_0 (1+ \epsilon \hat{\rho}), \quad
  \bu = \sqrt{\theta_0} \epsilon \hat{\bu}, \quad
  \theta = \theta_0 (1+ \epsilon \hat{\theta}), \\
& \bx = L \epsilon \hat{\bx}, \quad
  \tau = \frac{L}{\sqrt{\theta_0}} \epsilon \hat{\tau},
  \quad f_{\alpha} = \rho_0\theta_0^{|\alpha|/2} \epsilon \hat{f}_{\alpha}
    \quad \text{for} \quad |\alpha| \geqslant 1,
\end{split}
\end{equation}
where $\rho_0$ and $\theta_0$ are constants, $L$ is the characteristic
length, and variables with $\hat{\cdot}$ are dimensionless of
$O(1)$. Thus $\sigma_{ij}$ and $q_j$ can be expressed as
\begin{equation} \label{eq:sigma_q}
\sigma_{ij} = \rho_0 \theta_0 \epsilon \hat{\sigma}_{ij}, \qquad
q_j = \rho_0 \theta_0^{3/2} \epsilon \hat{q}_j.
\end{equation}
Now we substitute \eqref{eq:linearize} and \eqref{eq:sigma_q} for the
corresponding terms in \eqref{eq:reg_term}. After eliminating the
constant factors on both sides, the result reads
\begin{equation} \label{eq:hat_f_alpha}
\begin{split}
  \hat{f}_{\alpha} &= \hat{\tau} \left[
    \frac{1}{1 + \epsilon \hat{\rho}} \sum_{j=1}^D
      \pd{(1+\epsilon \hat{\rho})(1+\epsilon \hat{\theta})}{\hat{x}_j}
      \hat{f}_{\alpha-e_j}
    - \sum_{j=1}^D
      (1 + \epsilon \hat{\theta}) \pd{\hat{f}_{\alpha-e_j}}{\hat{x}_j}
  \right] \\
  & \qquad {} + \frac{1}{1 + \epsilon \hat{\rho}}
    \sum_{j=1}^D \sum_{d=1}^D \frac{1}{2} \epsilon^2
      \hat{\sigma}_{ij} \hat{f}_{\alpha-e_d-e_j} \\
  & \qquad {} + \frac{1}{1 + \epsilon \hat{\rho}}
    \sum_{j=1}^D \sum_{d=1}^D
      \frac{\epsilon^2\hat{q}_j}{(D+2)(1+\epsilon\hat{\theta})} \left[
        (1 + \epsilon \hat{\theta}) \hat{f}_{\alpha-2e_d-e_j} +
        (\alpha_j+1) \hat{f}_{\alpha-2e_d+e_j}
      \right].
\end{split}
\end{equation}
After collecting the terms of $O(\epsilon)$, \eqref{eq:hat_f_alpha}
is reformulated as
\begin{equation} \label{eq:linearized}
  \hat{f}_{\alpha} = -\hat{\tau} \sum_{j=1}^D
  (1 +  \epsilon \hat{\theta}) \pd{\hat{f}_{\alpha - e_j}}{\hat{x}_j}
  + O(\epsilon),
\end{equation}
and the $O(\epsilon)$ term is then simply dropped. Note that one
$O(\epsilon)$ term is intentionally kept in \eqref{eq:linearized}
for the convenience of variable restoration, which is performed by
\begin{equation} \label{eq:linear_reg_term}
\begin{split}
  f_{\alpha} & = \rho_0 \theta_0^{|\alpha| / 2} \epsilon
  \hat{f}_{\alpha} = -\rho_0 \theta_0^{|\alpha| / 2} \epsilon
  \hat{\tau} \sum_{j=1}^D
  (1 + \epsilon \hat{\theta}) \pd{\hat{f}_{\alpha-e_j}}{\hat{x}_j} \\
  & = -\frac{L}{\sqrt{\theta_0}} \epsilon \hat{\tau} \cdot \theta_0 (1
  + \epsilon \hat{\theta}) \sum_{j=1}^D \pd{(\rho_0
    \theta_0^{(|\alpha|-1)/2} \epsilon \hat{f}_{\alpha-e_j})} {(L
    \epsilon \hat{x}_j)} = -\tau \theta \sum_{j=1}^D \pd{f_{\alpha -
      e_j}}{x_j}.
\end{split}
\end{equation}
Obviously, \eqref{eq:linear_reg_term} is much neater than \eqref
{eq:reg_term}, and this linearized regularization term is used in our
numerical examples.

\begin{remark}
In the 1D case, the regularization term \eqref{eq:linear_reg_term}
becomes
\begin{equation} \label{eq:reg_term_1D}
f_{\alpha} = -\tau \theta \pd{f_{\alpha - e_1}}{x},
  \quad |\alpha| = M + 1.
\end{equation}
And this term is only used in the following term in
\eqref{eq:moment_eqs}
\begin{equation} \label{eq:diffuse_term_1D}
(\alpha_1 + 1) \pd{f_{\alpha + e_1}}{x}, \quad |\alpha| = M.
\end{equation}
The equations \eqref{eq:reg_term_1D} and \eqref{eq:diffuse_term_1D}
yields a diffusion on the $M$-th order term
\begin{equation}
-\pd{}{x} \left(
  (\alpha_1 + 1) \tau \theta \pd{f_{\alpha}}{x}
\right), \quad |\alpha| = M,
\end{equation}
which reveals the effect of regularization on the Grad-type systems.
\end{remark}

\begin{remark}
  In the case that $M$ is not a multiple of $3$, one may find that the
  linearized regularization term \eqref{eq:linear_reg_term} becomes
  $O(\tau^{\lceil |\alpha| / 3 \rceil + 1})$ while it ought to be
  $O(\tau^{\lceil |\alpha| / 3 \rceil})$. This is caused by using
  different conceptions of ``magnitude'' between regularization and
  linearization. Despite of this, the linear regularization \eqref
  {eq:linear_reg_term} indeed convert the moment system to a parabolic
  one. As known, Grad's moment equations restrict the distribution
  function in a pseudo-equilibrium manifold (see \cite{Struchtrup}),
  but the regularization \eqref{eq:linear_reg_term} relieves such
  restriction and allow a small perturbation around the manifold.
  Although the perturbation may not be large enough, it introduces
  additional flexibility for the moment system to agree with the real
  physics. Also, in our numerical experiments, only slight difference
  can be found between \eqref{eq:reg_term} and
  \eqref{eq:linear_reg_term}.
\end{remark}

\begin{remark} \label{rem:lin}
  One may argue that the large moment system is aimed at
  non-equilibrium fluids, and the assumption that the fluid is around
  a velocity-free equilibrium may lead to remarkable deviations.
  Actually, if we define
  \begin{equation}
  g(\bxi) = \sum_{|\alpha| \leqslant M-3} f_{\alpha}
    \mathcal{H}_{\theta,\alpha}(\bv),
  \end{equation}
  and then linearize \eqref{eq:reg_term} around $g(\bxi)$ instead of the
  Maxwellian, it can be found that the linearized result is exactly as
  \eqref{eq:linear_reg_term}. This explains why there is no
  significant difference between the linear and nonlinear
  regularizations in numerical results when $M$ is large.
\end{remark}

\subsection{Comparison with earlier approaches}
In \cite{NRxx}, an approach to solve moment system of arbitrary order
has been proposed. There we use the asymptotic expansion
\begin{equation} \label{eq:Chapman-Enskog}
f = f_0 + \varepsilon f_1 + \varepsilon^2 f_2 + \cdots
\end{equation}
to derive an approximation of $f_1$, where $f_0$ is the $M$-th order
Hermite expansion:
\begin{equation}
f_0 = \sum_{|\alpha| \leqslant M}
  f_{\alpha} \mathcal{H}_{\theta,\alpha} \left(
    \frac{\boldsymbol{\xi} - \boldsymbol{u}}{\sqrt{\theta}}
  \right).
\end{equation}
And the result is
\begin{equation} \label{eq:f1}
f_1 = -\tau \left(
  \pd{f_0}{t} + \bxi \cdot \nabla_{\bx} f_0
\right).
\end{equation}
The similarity between the method above and the current approach is
clear. Actually, \eqref{eq:iter_s} can be obtained by taking moments
on both sides of \eqref{eq:f1} and collecting some ``high order
terms''. Below we explain the differences between these two methods,
which are the our major motivation to write this paper.

The first difference comes from a defect in the theory of the earlier
method.  In \eqref{eq:Chapman-Enskog}, the part $f - f_0$ is scaled by
a small pseudo-timescale $\varepsilon$. However, it is not clear why
this part can be considered as ``small''. This is now clarified by the
order-of-magnitude approach since the magnitude of each moment has
been made clear.

As has been reported in Section \ref{sec:intro}, there exist some
circumstances when the earlier method fails due to the possible loss
of hyperbolicity while the new method does not. According to the
common theory of the Grad-type methods, this only happens when the
solution is relatively far away from Maxwellian, which means the
``high order terms'' that have been thrown away in this new approach
cannot be simply neglected. However, it is extremely complicated how
these terms affect the hyperbolicity of the equations, which we have
no idea to make clear so far. Only in our numerical experiments, we
find that dropping these terms alleviates the problem of
hyperbolicity. One may argue that the new method decreases the
accuracy of the approximation, while in our opinion, the loss of
accuracy can be compensated for by increasing the number of moments.

Moreover, technical differences originate in designing numerical
schemes for these two different approaches. In \cite{NRxx}, we used
\eqref{eq:f1} directly as the regularization term. In the
discretization of \eqref{eq:f1}, a direct temporal difference was used
to approximate the temporal derivative. This is hard to be extended to
the second order scheme. And for the part of spatial derivatives, it
is known now that only three orders of moments contribute to $f_1$,
which is exactly what we have done in the new method. While in the
earlier numerical scheme, the difference of the whole distribution
function was used for approximation, so that all the moments have 
contribution to this term. Let us demonstrate this point in detail in
the 1D case: the old scheme approximates $F = \xi_1 \nabla_x f_0$ on
the $j$-th grid as
\begin{equation}
F_j = \frac{\xi_1 f_{0,j+1} - \xi_1 f_{0,j-1}}{2\Delta x},
\end{equation}
Now, consider the $(M+1)$-st order moment of $F_j$:
\begin{equation}
\begin{split}
F_{j,\alpha} &= C_{\theta_j, \alpha} \int_{\mathbb{R}^D}
  \mathcal{H}_{\theta_j,\alpha}(\boldsymbol{v}_j) F_j(\boldsymbol{\xi})
    \exp(|\boldsymbol{v}_j^2|/2) \,\mathrm{d}\boldsymbol{v}_j \\
&= \frac{C_{\theta_j,\alpha}}{2\Delta x} \left(
  \int_{\mathbb{R}^D}
    \mathcal{H}_{\theta_j,\alpha}(\boldsymbol{v}_j) \left[
      \xi_1 f_{0,j+1}(\boldsymbol{\xi}) -
      \xi_1 f_{0,j-1}(\boldsymbol{\xi})
  \right] \exp(|\boldsymbol{v}_j^2|/2) \,\mathrm{d}\boldsymbol{v}_j
\right),
\end{split}
\end{equation}
where $\boldsymbol{v}_j = (\boldsymbol{\xi} - \boldsymbol{u}_j) /
\sqrt{\theta_j}$, $|\alpha| = M + 1$, and
\begin{equation}
C_{\theta_j,\alpha} =
  \frac{(2\pi)^{D/2} \theta_j^{|\alpha| + D}}{\alpha!}.
\end{equation}
Since $(\boldsymbol{u}_j, \theta_j) \neq (\boldsymbol{u}_{j-1},
\theta_{j-1})$ and $(\boldsymbol{u}_j, \theta_j) \neq
(\boldsymbol{u}_{j+1}, \theta_{j+1})$ in general, the above
calculation requires projections. Thus all moments of $f_{0,j+1}$ and
$f_{0,j-1}$ contributes to $F_{j,\alpha}$. In the new method, we first
write the $(M+1)$-st order moments of $F$ as
\begin{equation}
\begin{split}
F_{\alpha} &= C_{\theta,\alpha} \int_{\mathbb{R}^D}
  \mathcal{H}_{\theta,\alpha}(\boldsymbol{v}_j)
  \exp(|\boldsymbol{v}_j|^2 / 2) \cdot
  \xi_1 \nabla_{x} f_0 \,\mathrm{d} \boldsymbol{v}_j \\
&= \theta \frac{\partial f_{0,\alpha-e_1}}{\partial x}
  + \sum_{d=1}^D \frac{\partial u_d}{\partial x} \left(
    \theta f_{0,\alpha-e_d-e_1} + u_1 f_{0,\alpha-e_d}
  \right) \\
& \qquad + \frac{1}{2} \frac{\partial \theta}{\partial x}
  \sum_{d=1}^D (\theta f_{0,\alpha-2e_d-e_1} + u_1 f_{0,\alpha-2e_d}
    + (\alpha_1 + 1) f_{0,\alpha-2e_d+e_1}),
\end{split}
\end{equation}
and then this equation is adopted to design the numerical scheme. In
this expression, it is clear that only three orders of moments have
contribution to $F_{\alpha}$. This is the major difference between the
two methods in the numerical fold, which might lead to large
deviations in calculating numerical fluxes. The underlying reason of
this difference lies in different understandings of the regularization
term, although such disagreement can be eliminated by the refinement
of the computational mesh.

Additionally, the new scheme is more efficient than the earlier one.
Currently the most expensive part in the algorithm is the projection,
which requires to solve an ordinary differential system. The
above-mentioned differencing of distributions requires twice of such
projection, which is no longer needed in the new framework.
Considering the construction of a first order numerical scheme, it is
found that the computational time can be almost halved by such
improvement.

%%% Local Variables: 
%%% mode: latex
%%% TeX-master: "article"
%%% End: 

\section{Numerical examples}

In this section, two numerical examples of our method for
the regularized moment equations are presented. In both
tests, the global Knudsen number is denoted as $\Kn$, and
The CFL number is always $0.95$. We use the POSIX multi-threaded
technique in our simulation, and at most $8$ CPU cores are used.

\subsection{Shock tube test}

To demonstrate that the new method is applicable to the
examples in \cite{NRxx}, the first example is a repetition
of the shock tube problem in \cite{NRxx, Torrilhon2006}.

As in \cite{NRxx, Torrilhon2006}, the initial conditions are
\begin{equation} 
\rho(0, x) = \begin{cases} 7.0, & x < 0, \\
1.0, & x > 0,
\end{cases} \qquad p(0, x) = \rho\theta = \begin{cases} 7.0,
& x < 0, \\ 1.0, & x > 0,
\end{cases} \qquad \bu = 0.
\end{equation}
The computational domain is $[-1, 1.5]$ and the stop time is $t =
0.3$. The relaxation time $\tau$ (see \eqref{eq:BGK}) is chosen as
$\Kn / \rho(t, \bu)$.  The numerical scheme is an improved version
\cite{Cai} of the method used in \cite{NRxx}, with the regularization
term substituted by \eqref{eq:linear_reg_term}. The improved numerical
scheme significantly reduces the computational cost by a large time
step method and high spatial resolution. Since the BGK model fails to
predict the correct Prandtl number, we plot the temperature instead of
the heat flux below.

\begin{figure}[ht] \centering
\begin{overpic}[scale=.45]{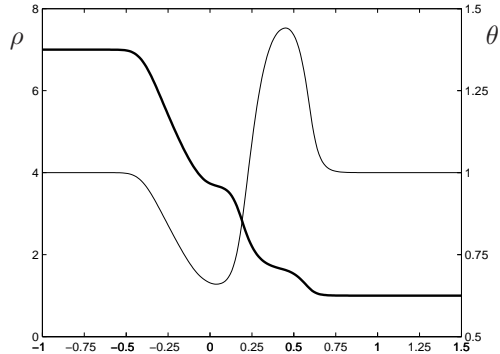}
\put(7,63){\small$\rho$}
\put(95,63){\small$\theta$}
\end{overpic}
\caption{Results for the shock tube test with $M = 3$ and
$\Kn = 0.02$. The thick curve with the left $y$-axis is the
plot of density. The thin curve with the right $y$-axis is
the plot of temperature. 200 grids are used in computation.}
\label{fig:kn=0.02}
\end{figure}

To validate the method with the new regularization term, we
compare the results produced by the new method with the
results in \cite{NRxx, Torrilhon2006} for both small and big
Knudsen numbers. For small Knudsen number $\Kn = 0.02$, the
plot of density for the distribution functions when $M = 3$
and $t =0.3$ is presented in Figure \ref{fig:kn=0.02}. The
density profile agrees with the result in \cite{NRxx,
  Torrilhon2006} perfectly.

For Knudsen number as great as $\Kn = 0.5$, both linearized
and non-linearized results from $M = 4$ to $M = 15$ are
computed, which are plotted in Figure \ref{fig:kn=0.5}.
Meanwhile, we solve the Boltzmann-BGK equation directly
using Mieussens' discrete velocity method \cite{Mieussens}
on a very fine mesh grid. One can find that the differences
between linearized and the original non-linear results are
only observable when $M$ is small, which verifies the
comments in Remark \ref{rem:lin}. Furthermore, the
computational results still converge to the BGK solution
gradually for both density and temperature, although the
convergence rate becomes much slower, as can be seen in
Figure \ref{fig:kn=0.5}.

\begin{figure}[!ht]
\centering
\subfigure[$M=4$]{
\begin{overpic}[width=.45\textwidth]{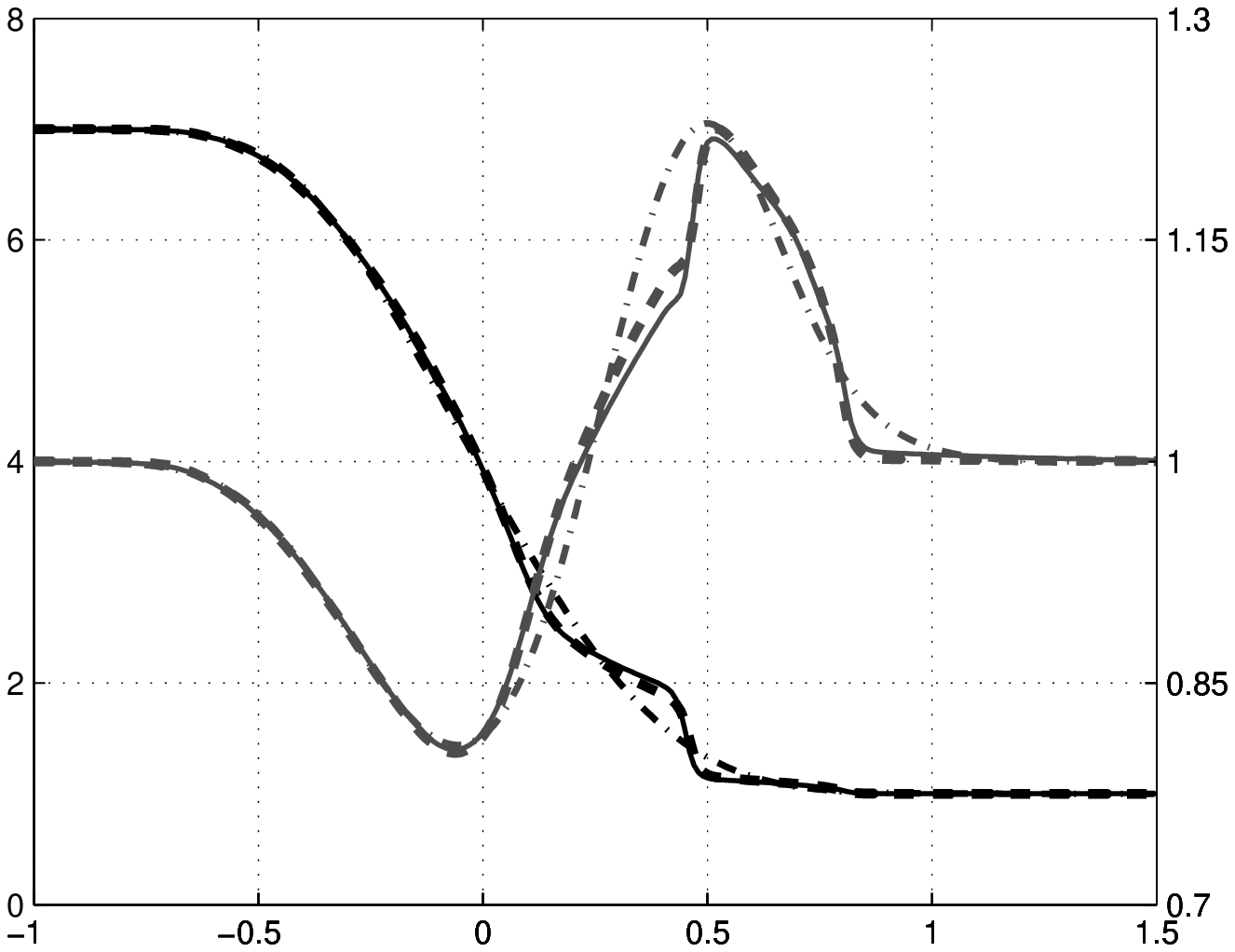}
\put(7,63){\small$\rho$}
\put(95,63){\small$\theta$}
\end{overpic}
}
\subfigure[$M=5$]{
\begin{overpic}[width=.45\textwidth]{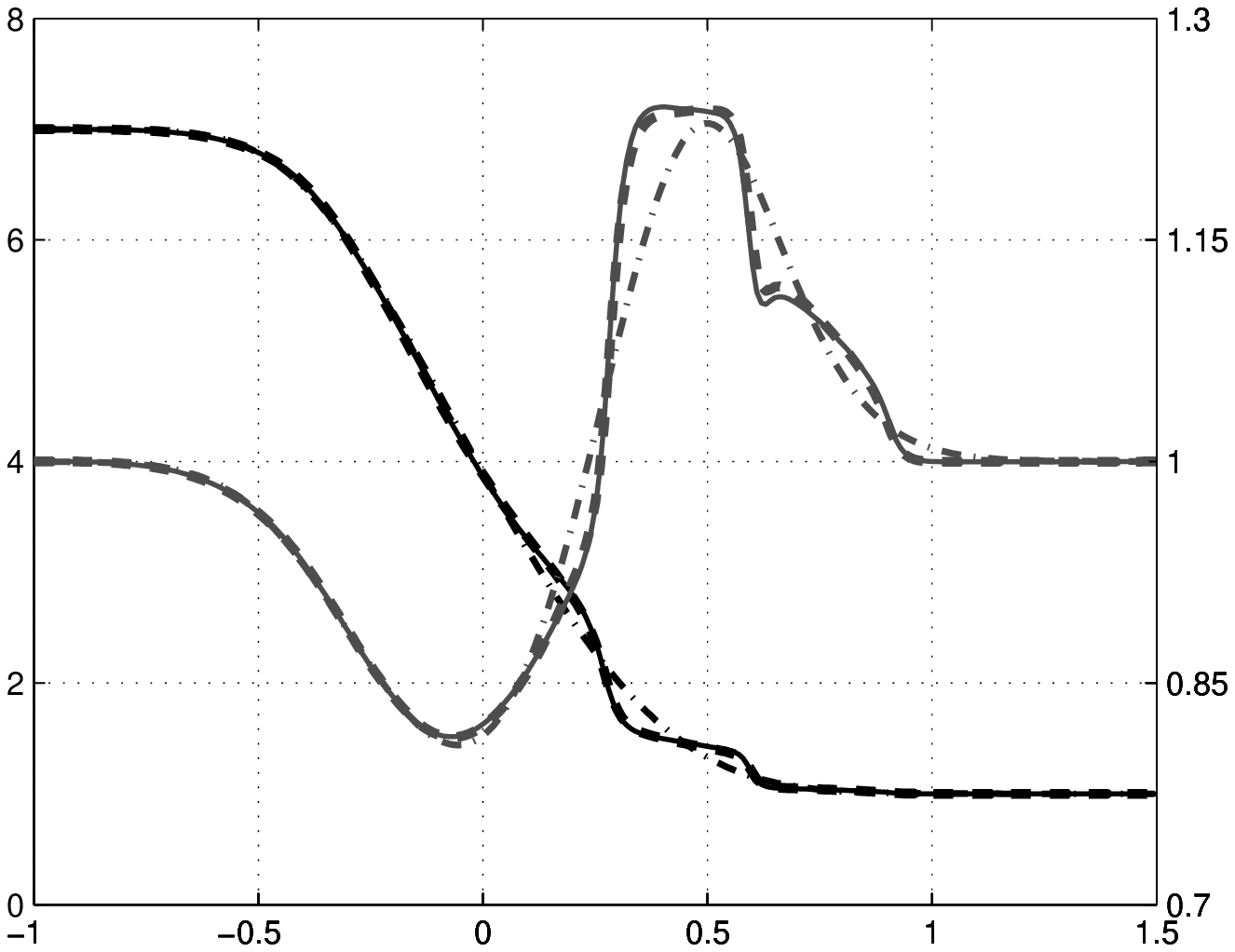}
\put(7,63){\small$\rho$}
\put(95,63){\small$\theta$}
\end{overpic}
}\\
\subfigure[$M=6$]{
\begin{overpic}[width=.45\textwidth]{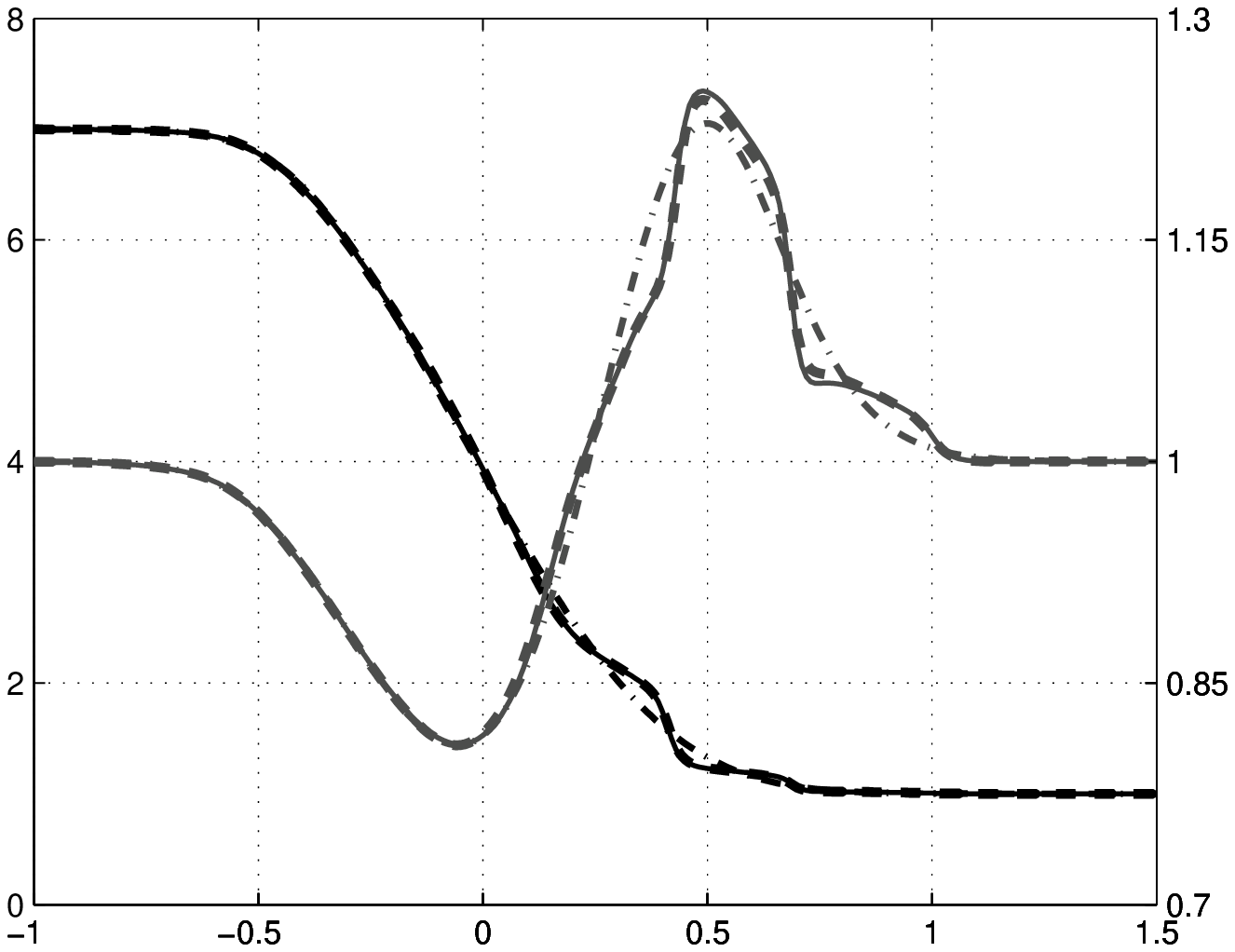}
\put(7,63){\small$\rho$}
\put(95,63){\small$\theta$}
\end{overpic}
}
\subfigure[$M=7$]{
\begin{overpic}[width=.45\textwidth]{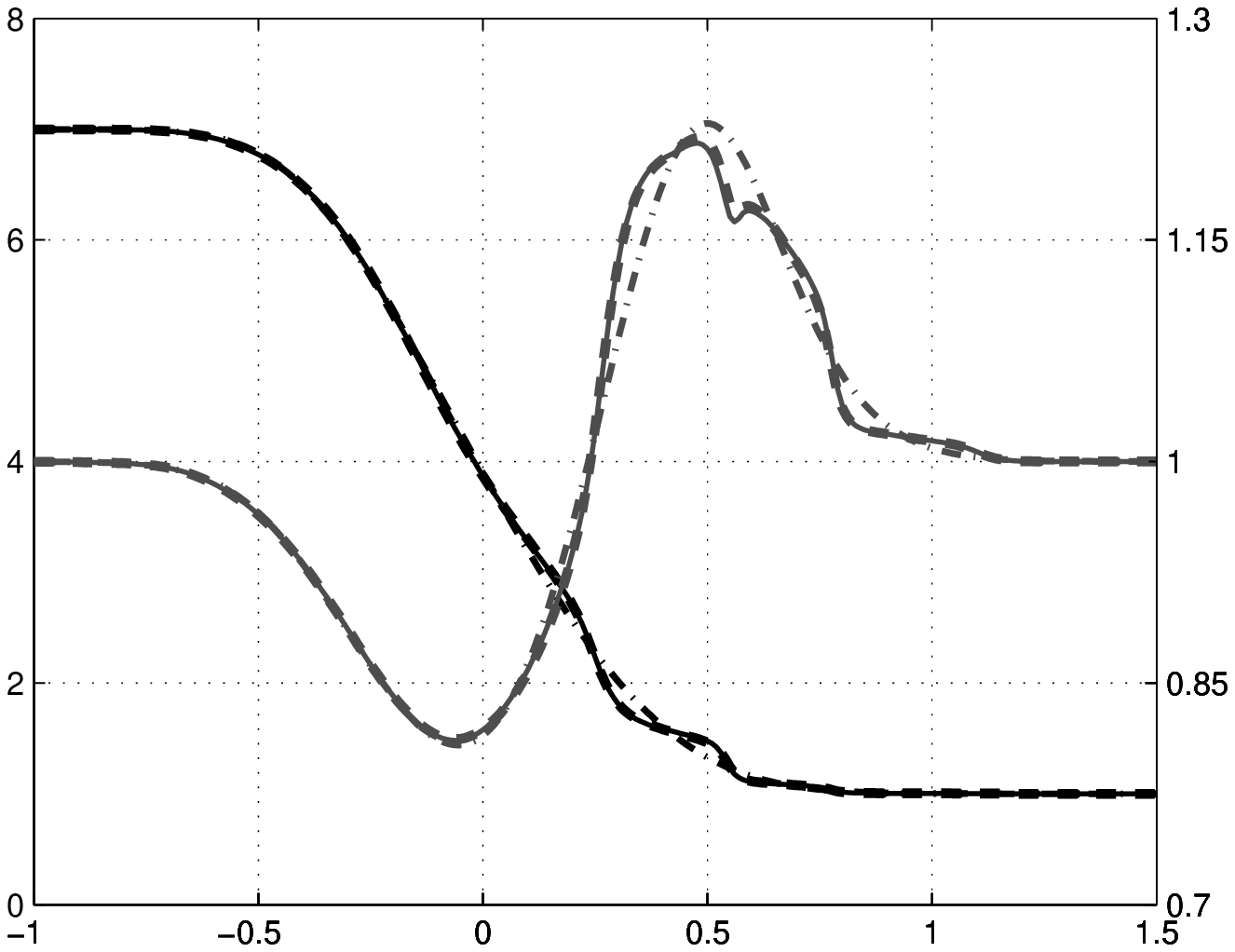}
\put(7,63){\small$\rho$}
\put(95,63){\small$\theta$}
\end{overpic}
}\\
\subfigure[$M=8$]{
\begin{overpic}[width=.45\textwidth]{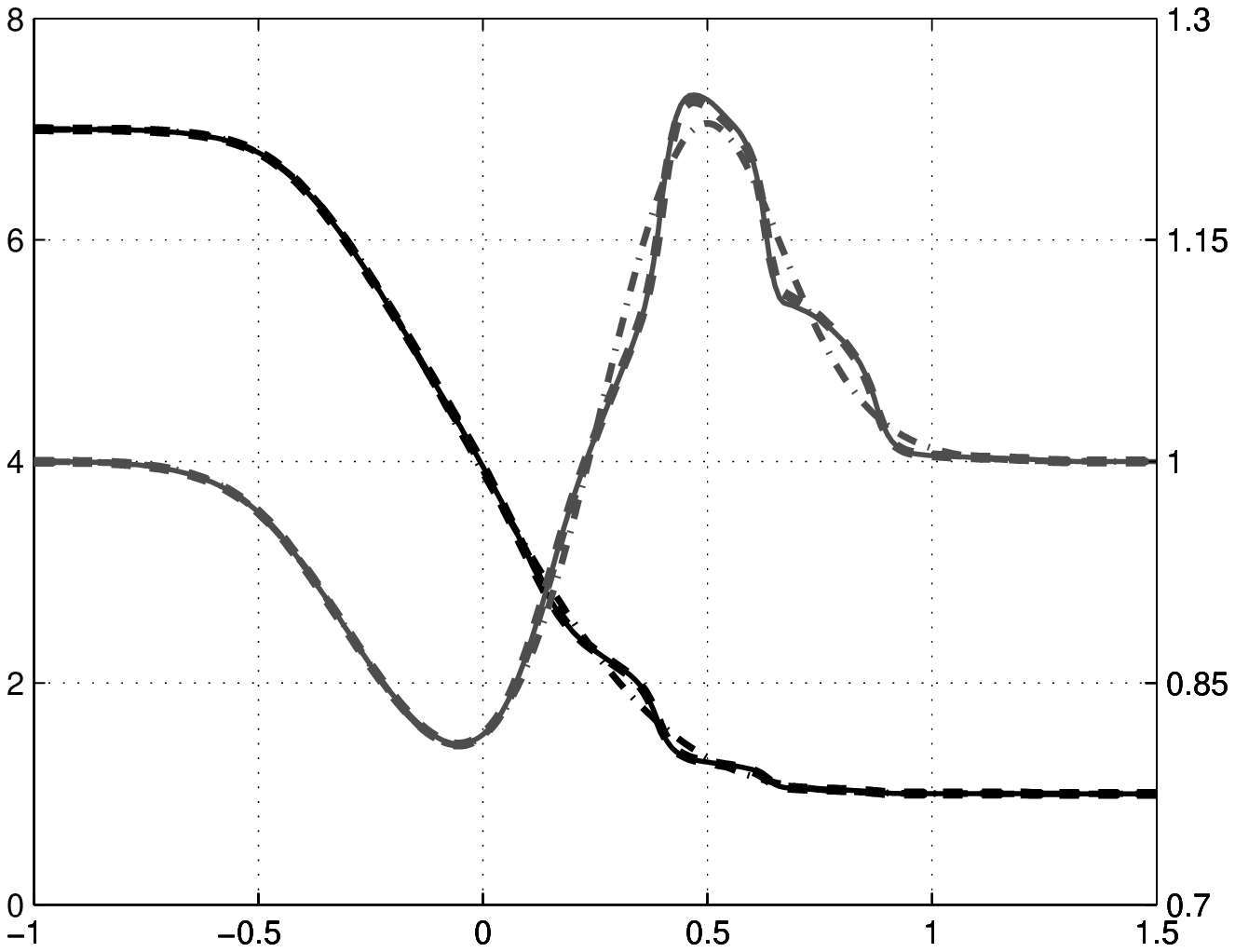}
\put(7,63){\small$\rho$}
\put(95,63){\small$\theta$}
\end{overpic}
}
\subfigure[$M=9$]{
\begin{overpic}[width=.45\textwidth]{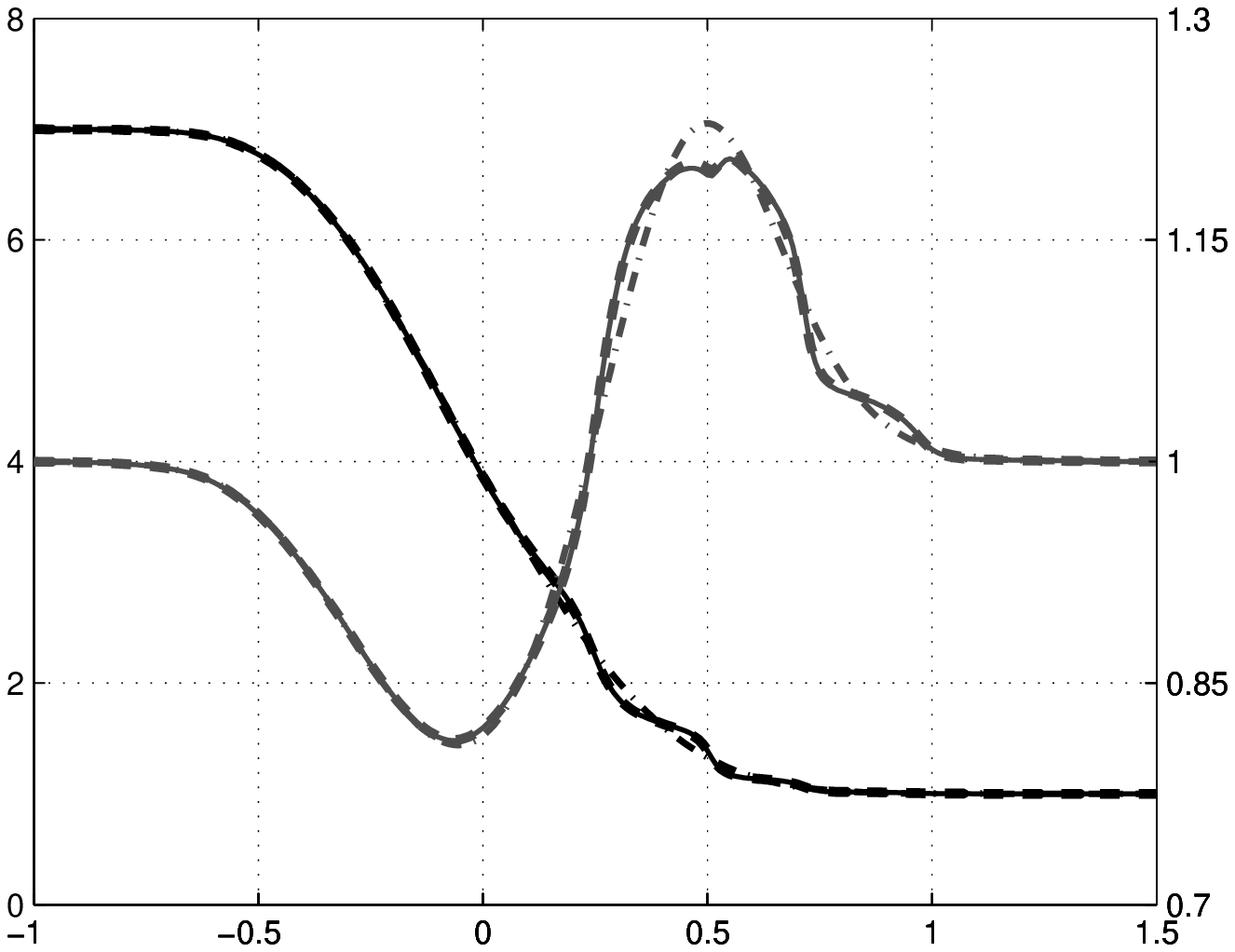}
\put(7,63){\small$\rho$}
\put(95,63){\small$\theta$}
\end{overpic}
}
\caption{Results for the shock tube test with $\Kn=0.5$. The dashed
lines are the results for regularized moment equations without
linearization, and the solid thin lines are the results with
linearization. The dashdot lines are the results of discrete velocity
model. The black lines denote the density and the gray lines denote
the temperature (to be continued).}
\end{figure}

\addtocounter{figure}{-1}

\begin{figure}[!ht]
\centering
\setcounter{subfigure}{6}
\subfigure[$M=10$]{
\begin{overpic}[width=.45\textwidth]{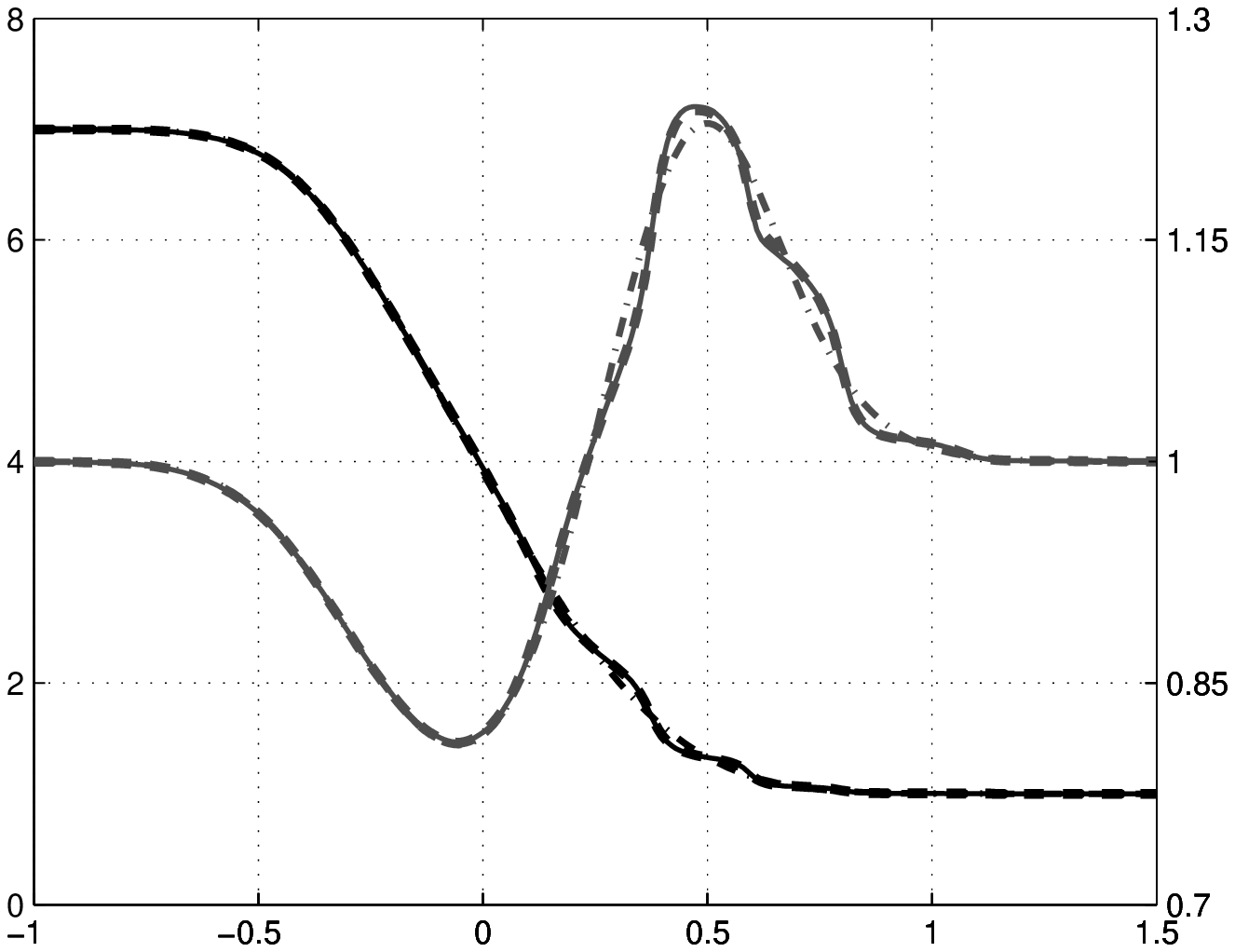}
\put(7,63){\small$\rho$}
\put(95,63){\small$\theta$}
\end{overpic}
}
\subfigure[$M=11$]{
\begin{overpic}[width=.45\textwidth]{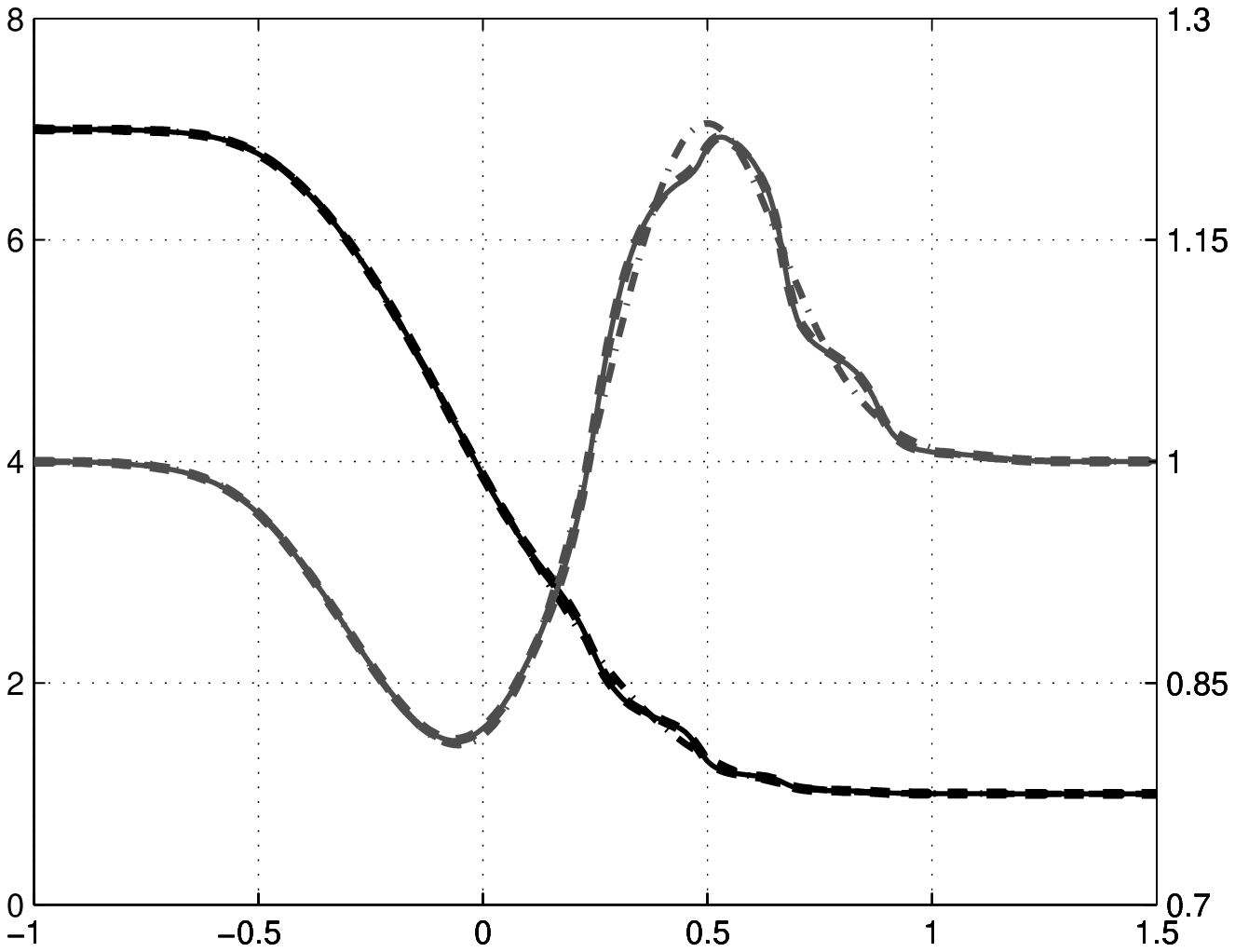}
\put(7,63){\small$\rho$}
\put(95,63){\small$\theta$}
\end{overpic}
}\\
\subfigure[$M=12$]{
\begin{overpic}[width=.45\textwidth]{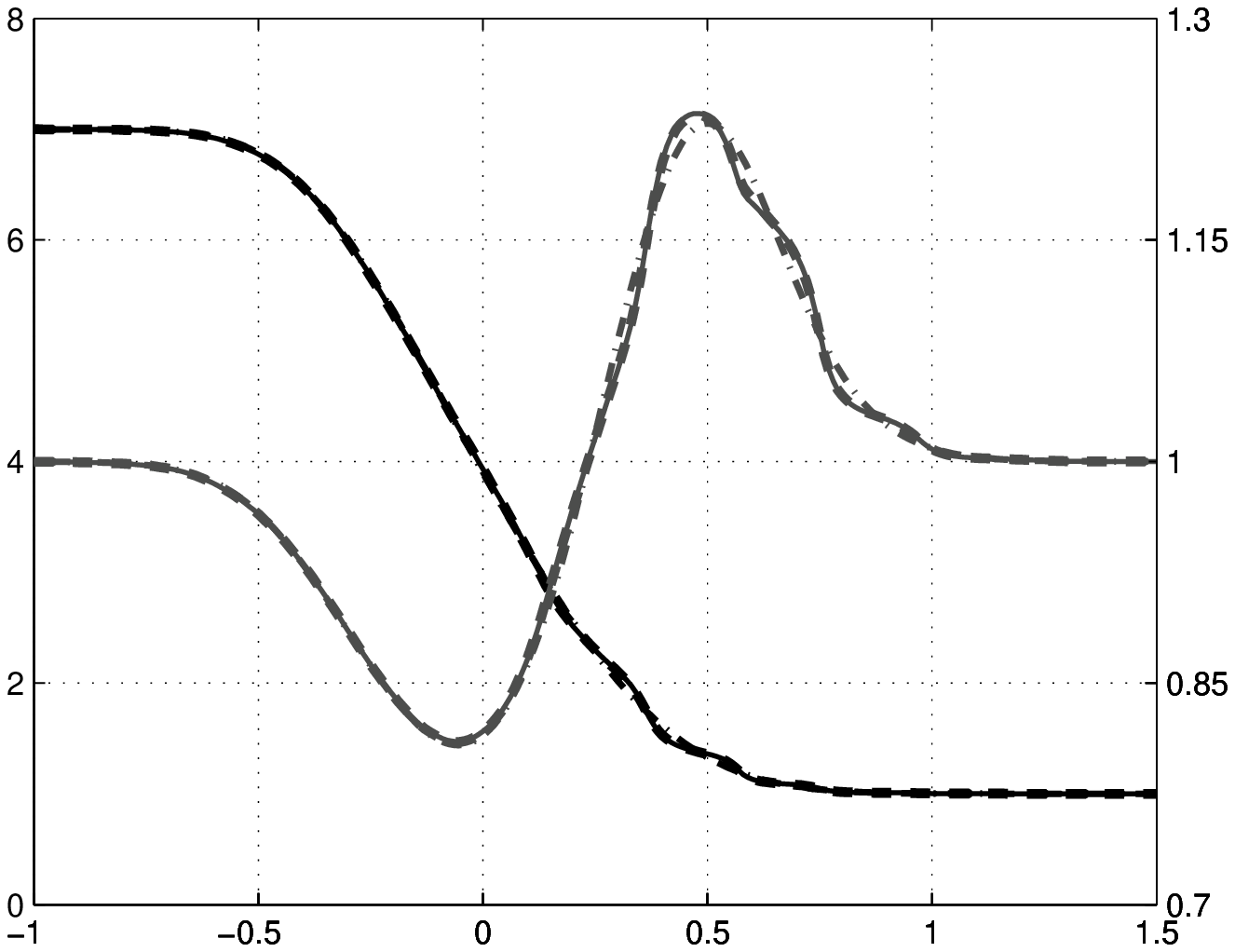}
\put(7,63){\small$\rho$}
\put(95,63){\small$\theta$}
\end{overpic}
}
\subfigure[$M=13$]{
\begin{overpic}[width=.45\textwidth]{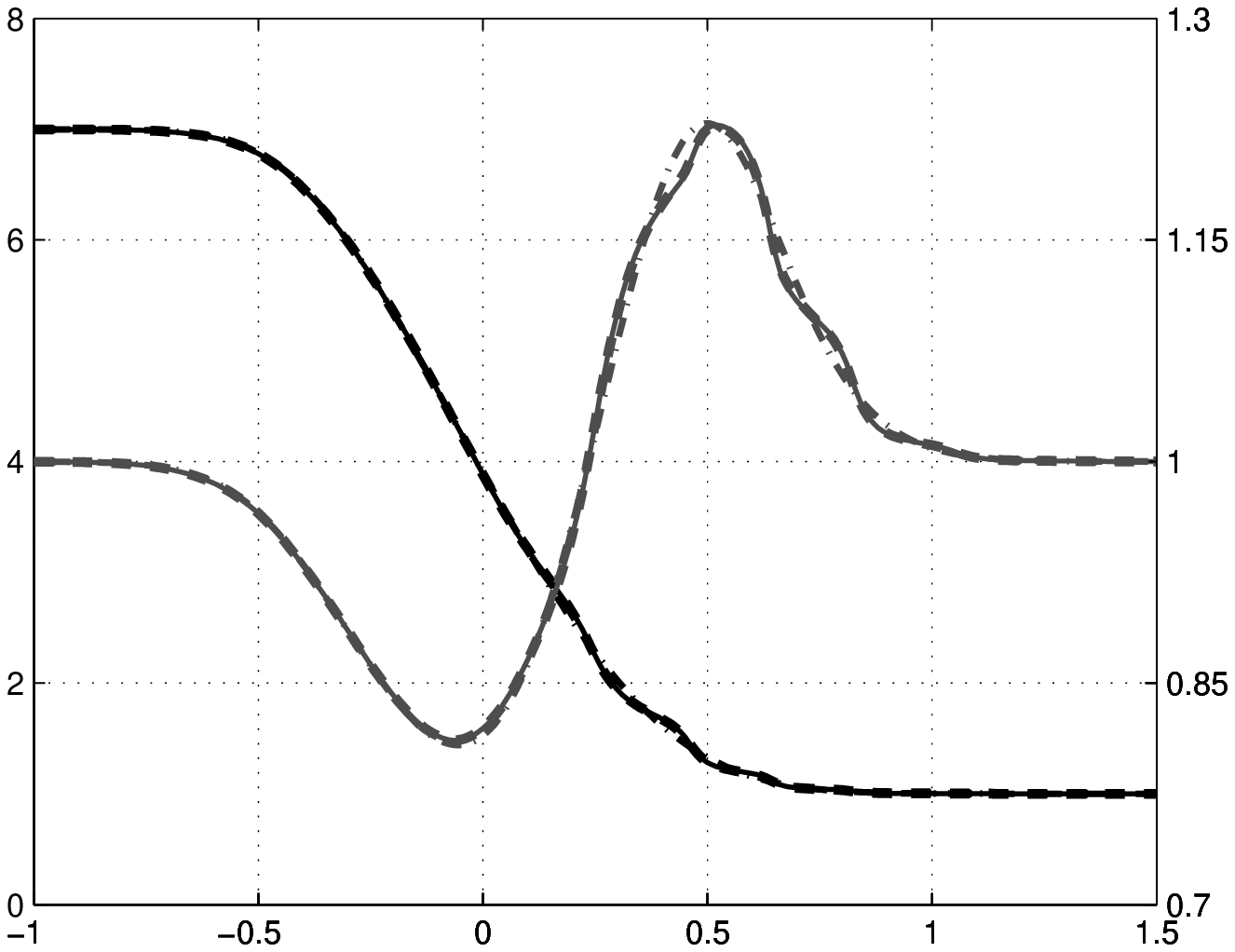}
\put(7,63){\small$\rho$}
\put(95,63){\small$\theta$}
\end{overpic}
}\\
\subfigure[$M=14$]{
\begin{overpic}[width=.45\textwidth]{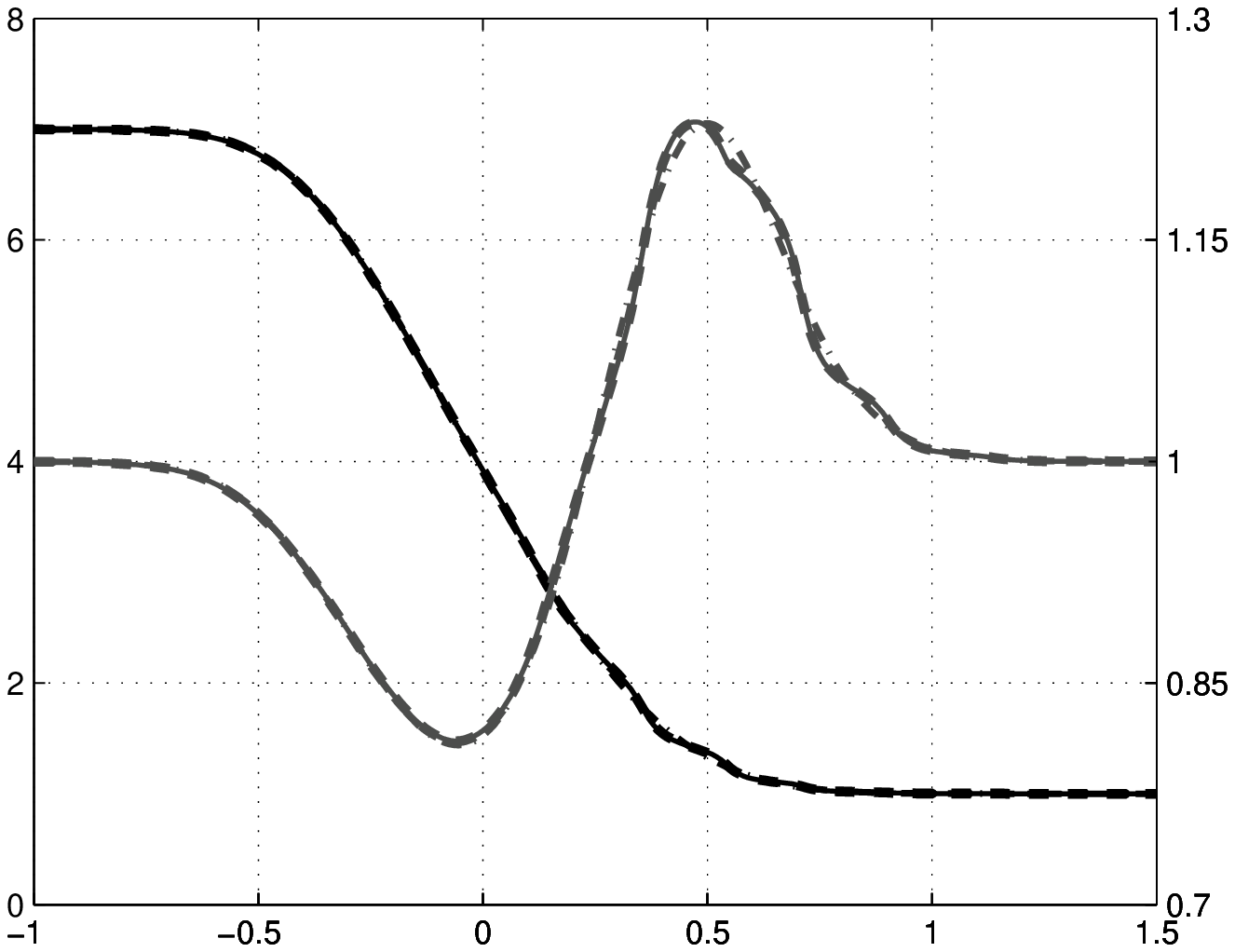}
\put(7,63){\small$\rho$}
\put(95,63){\small$\theta$}
\end{overpic}
}
\subfigure[$M=15$]{
\begin{overpic}[width=.45\textwidth]{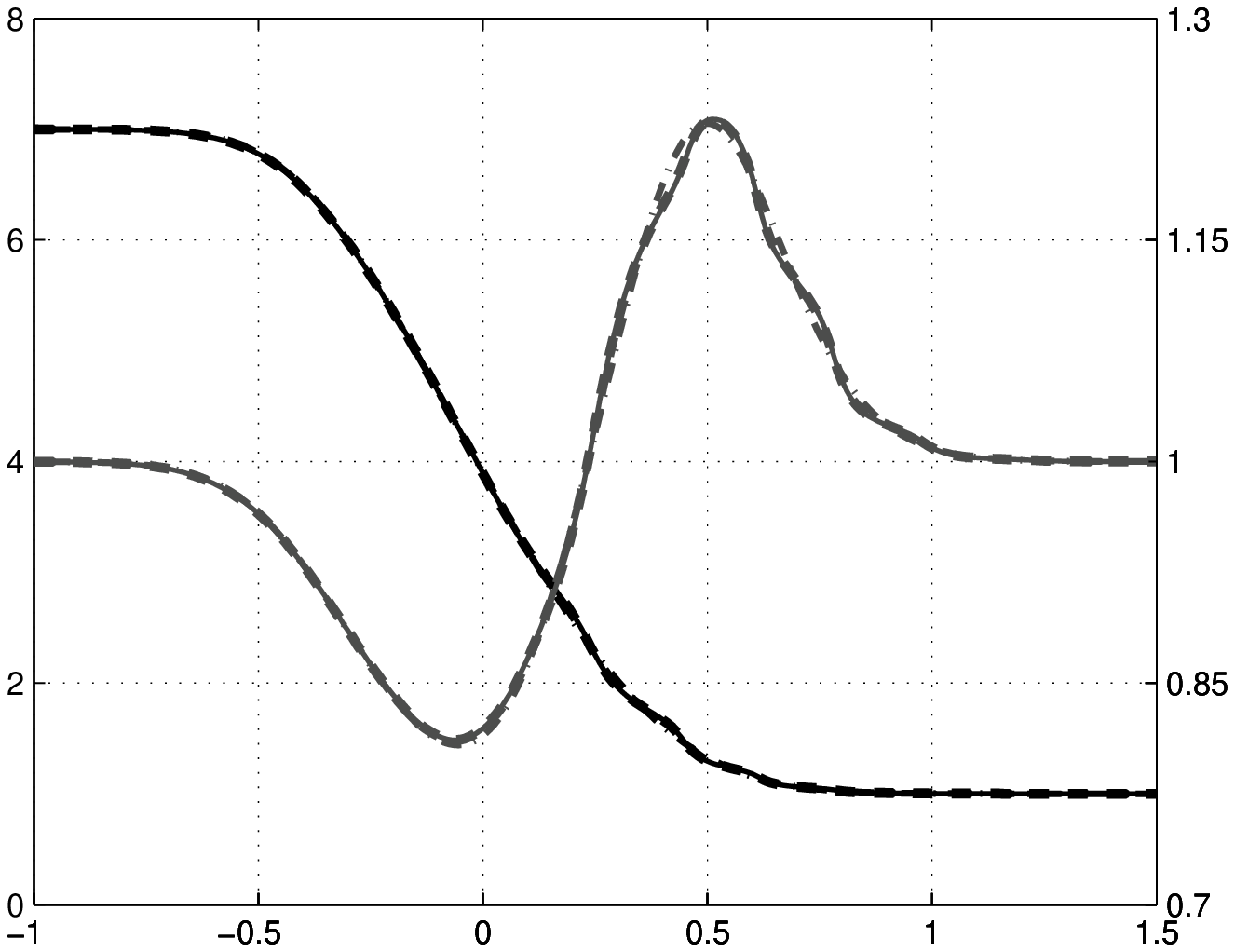}
\put(7,63){\small$\rho$}
\put(95,63){\small$\theta$}
\end{overpic}
}
\caption{Results for the shock tube test with $\Kn=0.5$. The dashed
lines are the results for regularized moment equations without
linearization, and the solid thin lines are the results with
linearization. The dashdot lines are the results of discrete velocity
model. The black lines denote the density and the gray lines denote
the temperature.}
\label{fig:kn=0.5}
\end{figure}

\begin{figure}
\centering
\subfigure[$M=3$]{
\begin{overpic}[width=.45\textwidth]{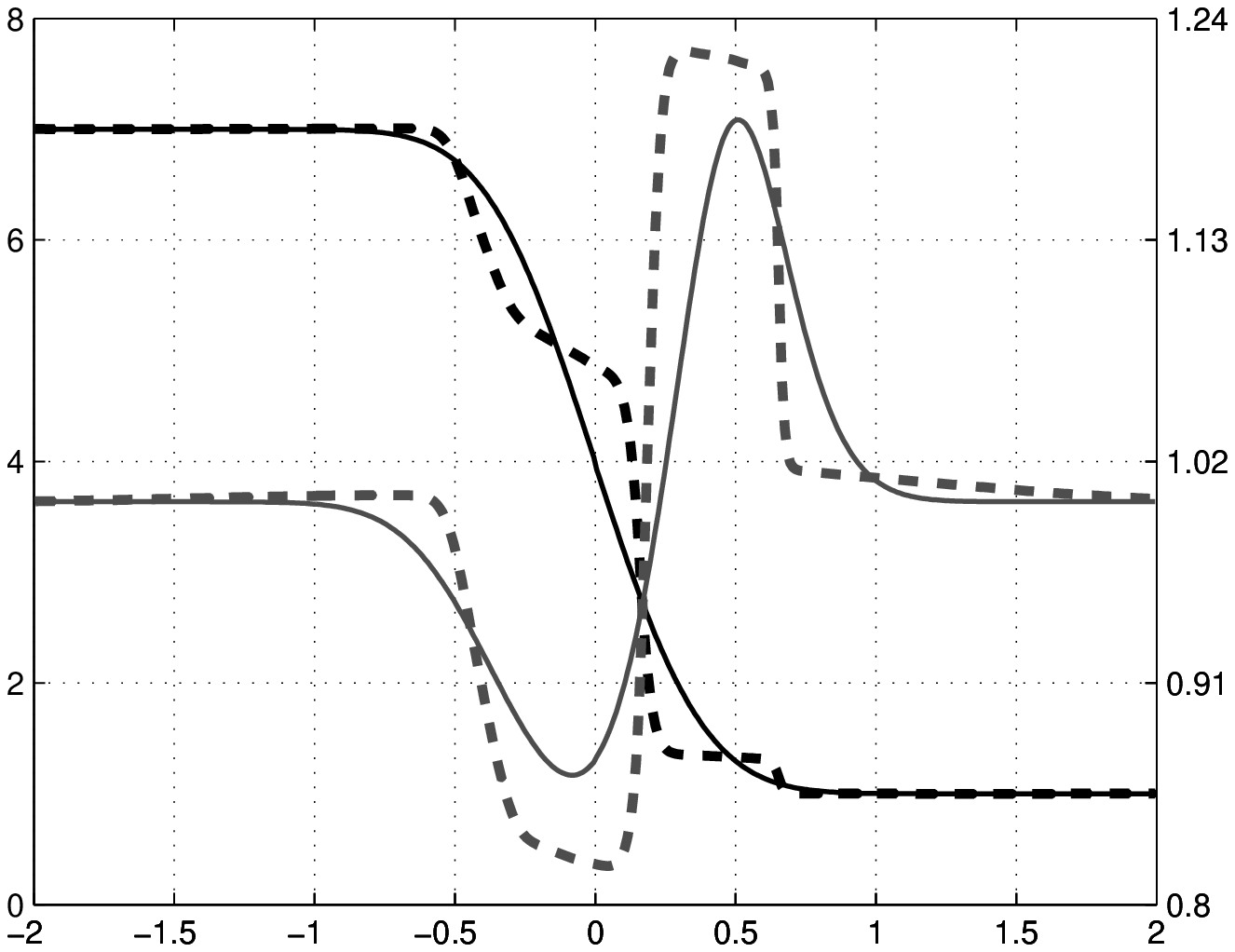}
\put(7,63){\small$\rho$}
\put(95,63){\small$\theta$}
\end{overpic}
}
\subfigure[$M=6$]{
\begin{overpic}[width=.45\textwidth]{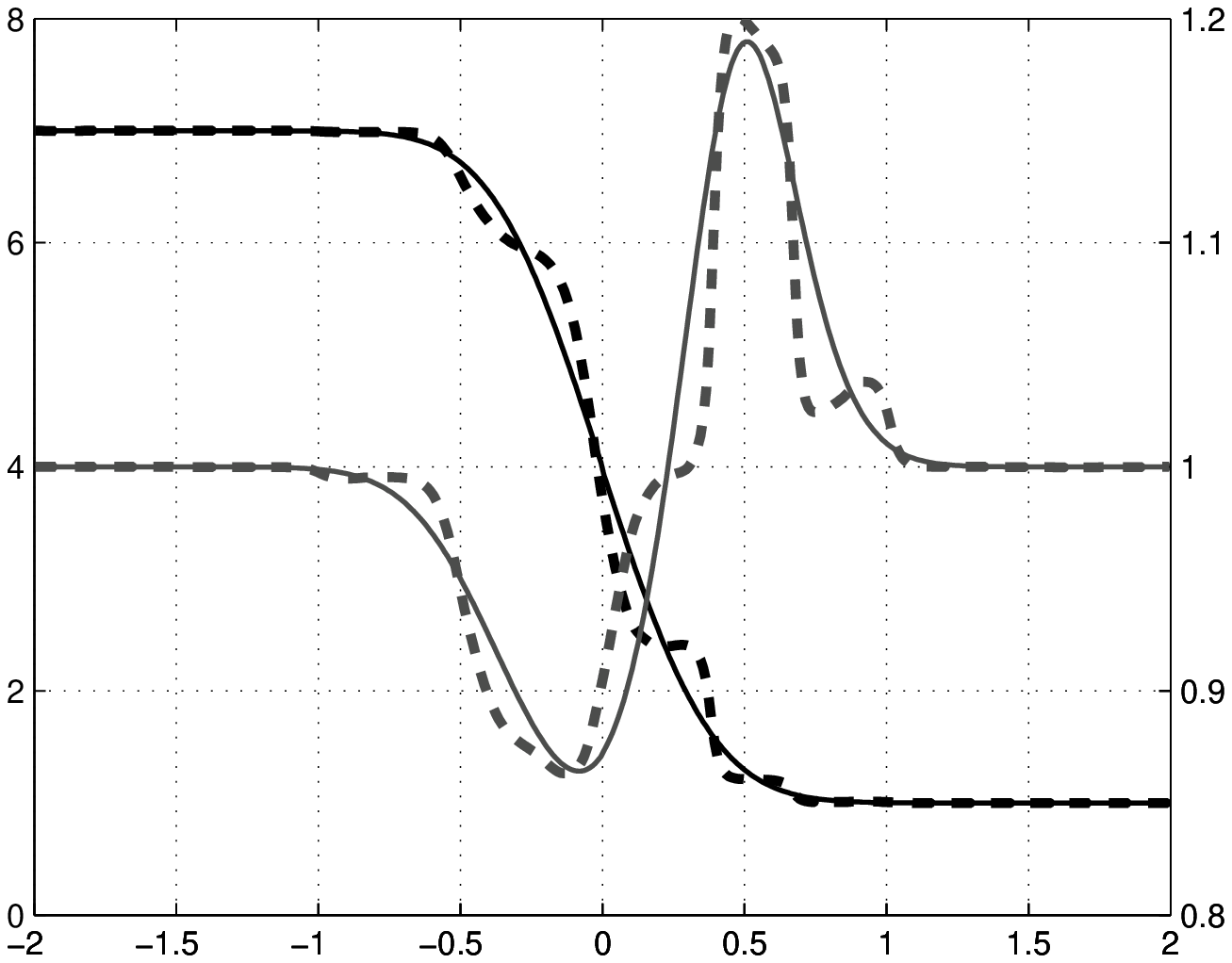}
\put(7,63){\small$\rho$}
\put(95,63){\small$\theta$}
\end{overpic}
}\\
\subfigure[$M=9$]{
\begin{overpic}[width=.45\textwidth]{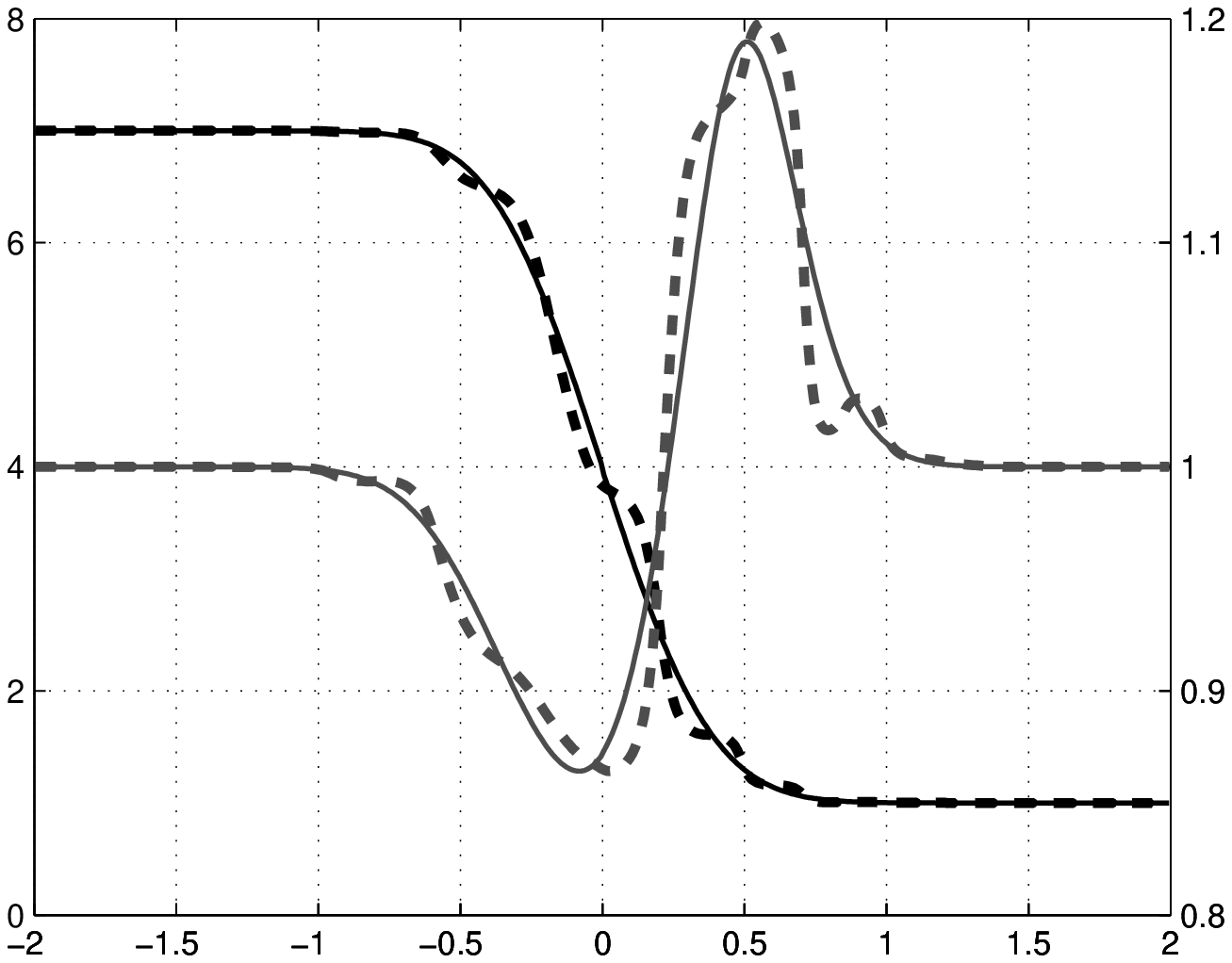}
\put(7,63){\small$\rho$}
\put(95,63){\small$\theta$}
\end{overpic}
}
\subfigure[$M=12$]{
\begin{overpic}[width=.45\textwidth]{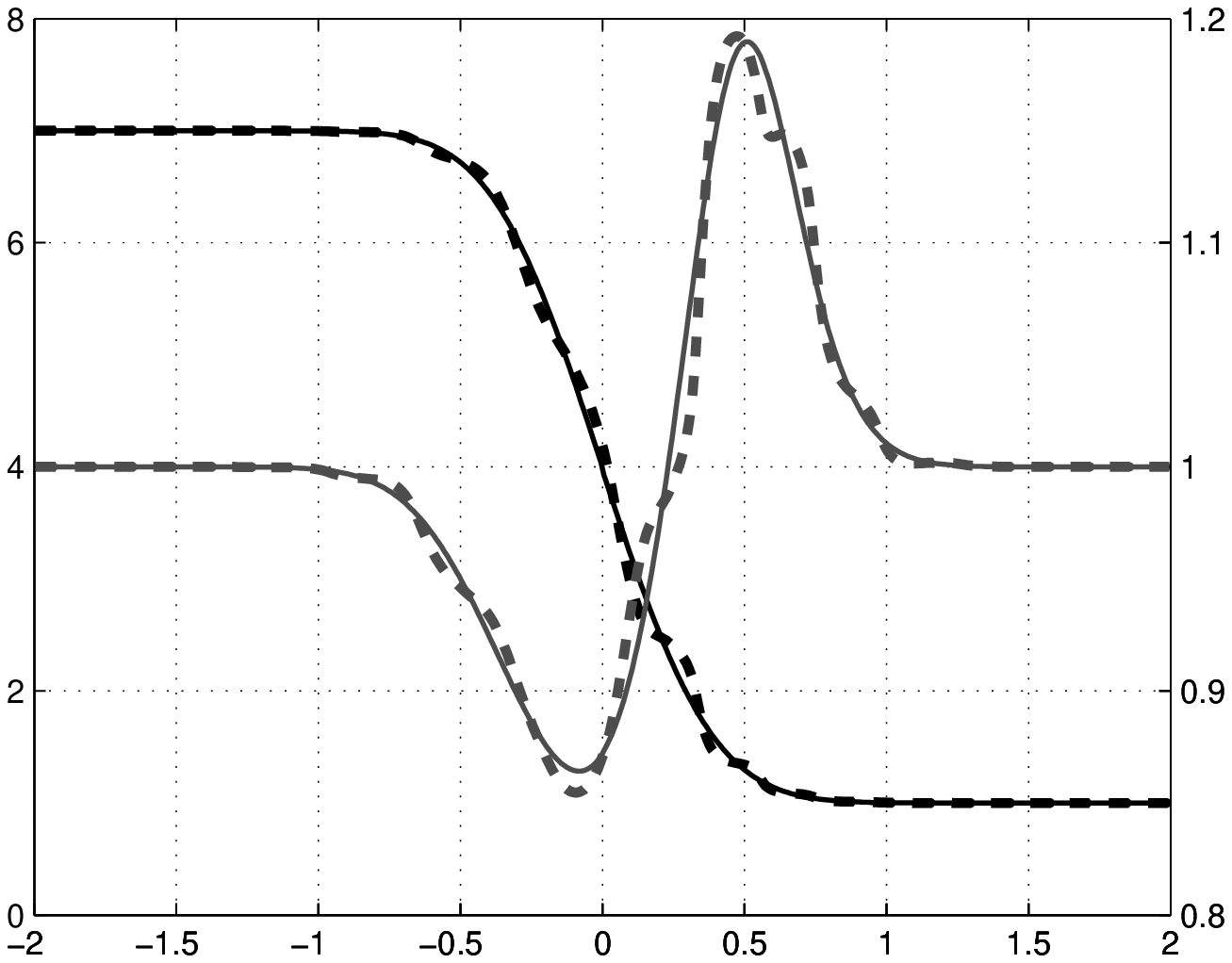}
\put(7,63){\small$\rho$}
\put(95,63){\small$\theta$}
\end{overpic}
}\\
\subfigure[$M=15$]{
\begin{overpic}[width=.45\textwidth]{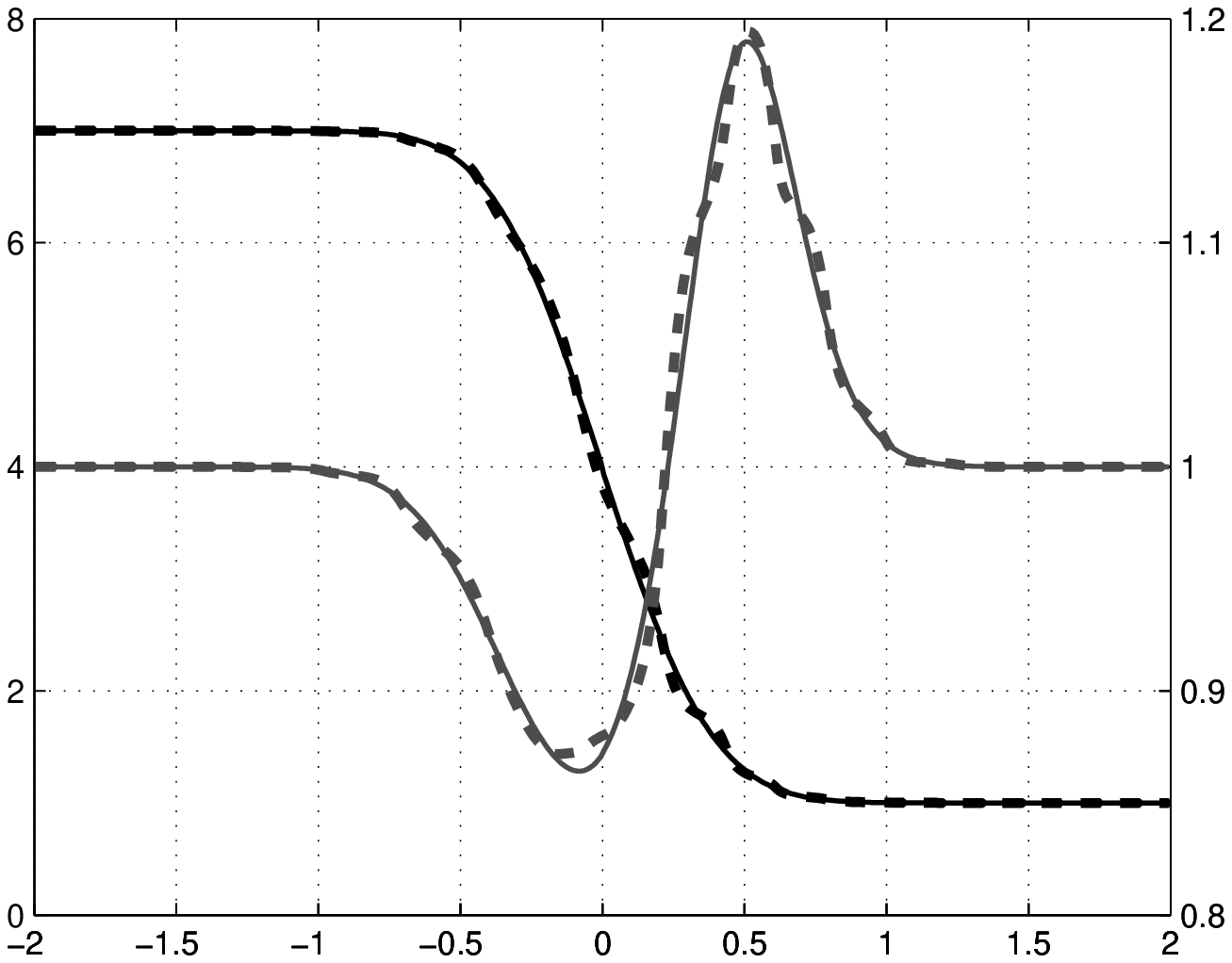}
\put(7,63){\small$\rho$}
\put(95,63){\small$\theta$}
\end{overpic}
}
\subfigure[$M=18$]{
\begin{overpic}[width=.45\textwidth]{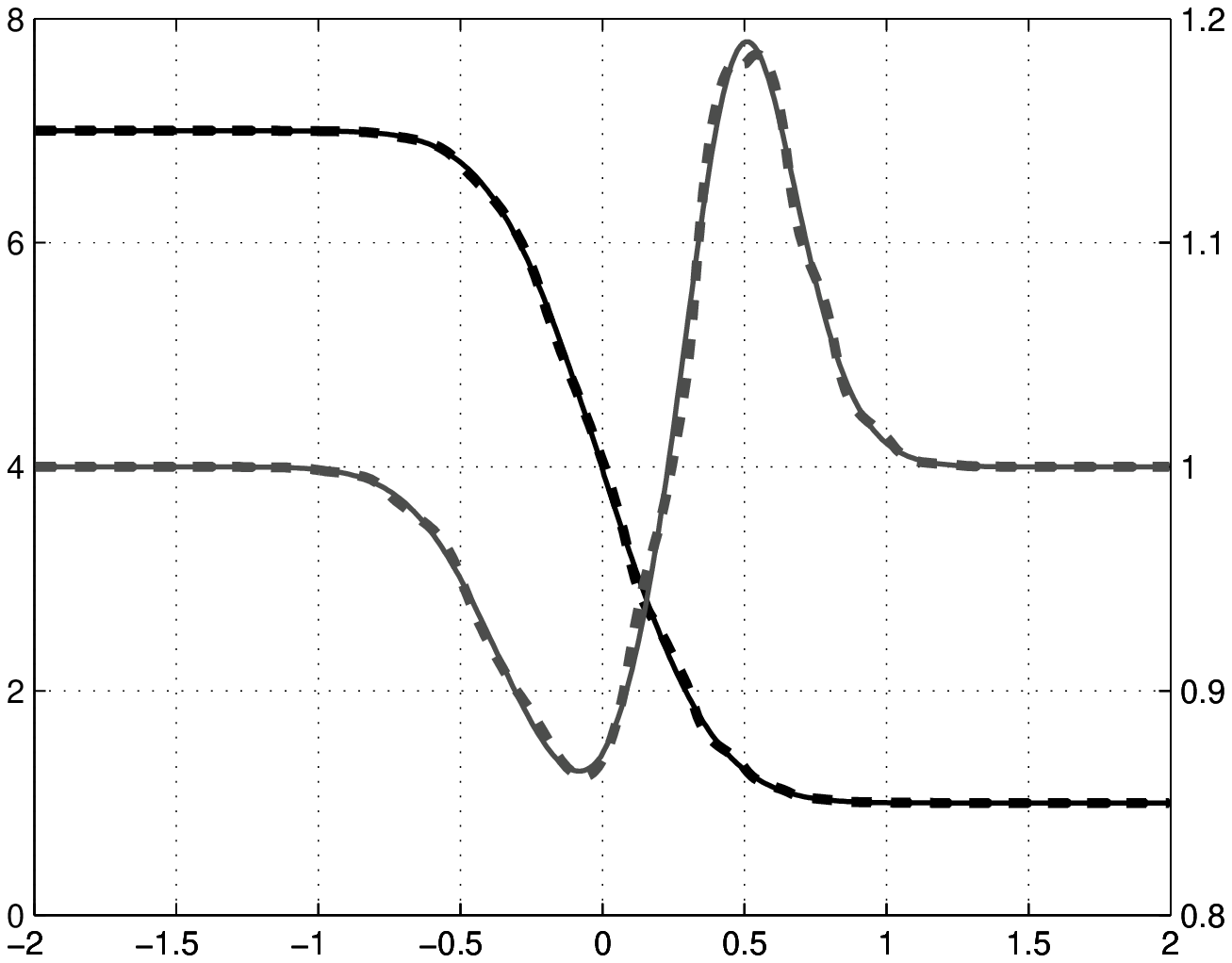}
\put(7,63){\small$\rho$}
\put(95,63){\small$\theta$}
\end{overpic}
}
\caption{Numerical results for shock tube problem with $\Kn=5$. The
dashed lines are the results for regularized moment equations with
linearization, and the solid thin lines are the results of the
discrete velocity model. The black lines denote the density and the
gray lines denote the temperature.}
\label{fig:kn=5}
\end{figure}

In \cite{Torrilhon_CiC}, it is pointed out that the
Grad-type moment system is not globally hyperbolic even for
13-moment system. In the case of a large ratio of the
density and the pressure, the solution leaves the region of
hyperbolicity, which leads to strong oscillations and
finally a breakdown of the computation \cite{Torrilhon_CiC}.
Though the ratio of the density and the pressure is as large
as $7.0$ in Figure \ref{fig:kn=0.5}, our method still works
well and produces converging results. In order to verify
this point, an even larger Knudsen number $\Kn = 5$ is
investigated. Results with $M=3,6,9,12,15,18$ are
considered, and the density and temperature profiles are
listed in Figure \ref{fig:kn=5}. Until $M=18$, the
regularzied moment method has given a satisfying agreement
with the discrete velocity model.

\subsection{Shock Structure Problem}
In this section, we carry out the simulation of a steady
shock structure with large Mach number. The shock structure
can be obtained by solving a 1D Riemann problem based on the
Rankine-Hugoniot condition. The left state is
\begin{equation} \label{eq:left_side}
\rho_r = 1, \quad u_r = \sqrt{\frac{5}{3}}M_0, \quad p_r = 1.
\end{equation}
and the right state is
\begin{equation} \label{eq:right_side}
\rho_r = \frac{4M_0^2}{M_0^2+3}, \quad
  u_r = \sqrt{\frac{5}{3}}\frac{M_0^2+3}{4M_0}, \quad
  p_r = \frac{5M_0^2 - 1}{4},
\end{equation}
Both states are in equilibrium. After a sufficiently long time, a
steady shock wave with fully developed structure can be obtained. This
example is aimed at the validation of our algorithm in high Mach
number. For high order moment systems, the current scheme still
suffers the problem of hyperbolicity. However, the R20 equations
($M=3$) turn out to be very robust in our numerical experiments. Here
we simulate the R20 equations with $M_0 = 1.53, 1.76, 2.05, 2.31,
3.38, 3.8, 6.5, 9.0$ and compare the results with the experimental
data in \cite{Alsmeyer}. The relaxation time is chosen as
\begin{equation}
\tau = \sqrt{\frac{\pi}{2}}
  \frac{15\Kn}{(5-2\omega)(7-2\omega)}
  \frac{\theta^{\omega-1}}{\rho},
\end{equation}
which is the result of the VHS model (see e.g. \cite{Bird}). The
constant $\omega$ is chosen as $0.72$ as suggested in \cite{Alsmeyer}.
The Knudsen number $\Kn = 1.0$ and spatial grid size $\Delta x = 0.1$
are used in this example. The results for the discrete velocity model
are also presented as a reference. All the plots are shown in Figure
\ref{fig:Shock-Structure}. The density has been normalized to the
interval $[0,1]$.

\begin{figure}[!ht] 
\centering
\subfigure[$M_0 = 1.55$]{
   \includegraphics[scale=.45,clip]{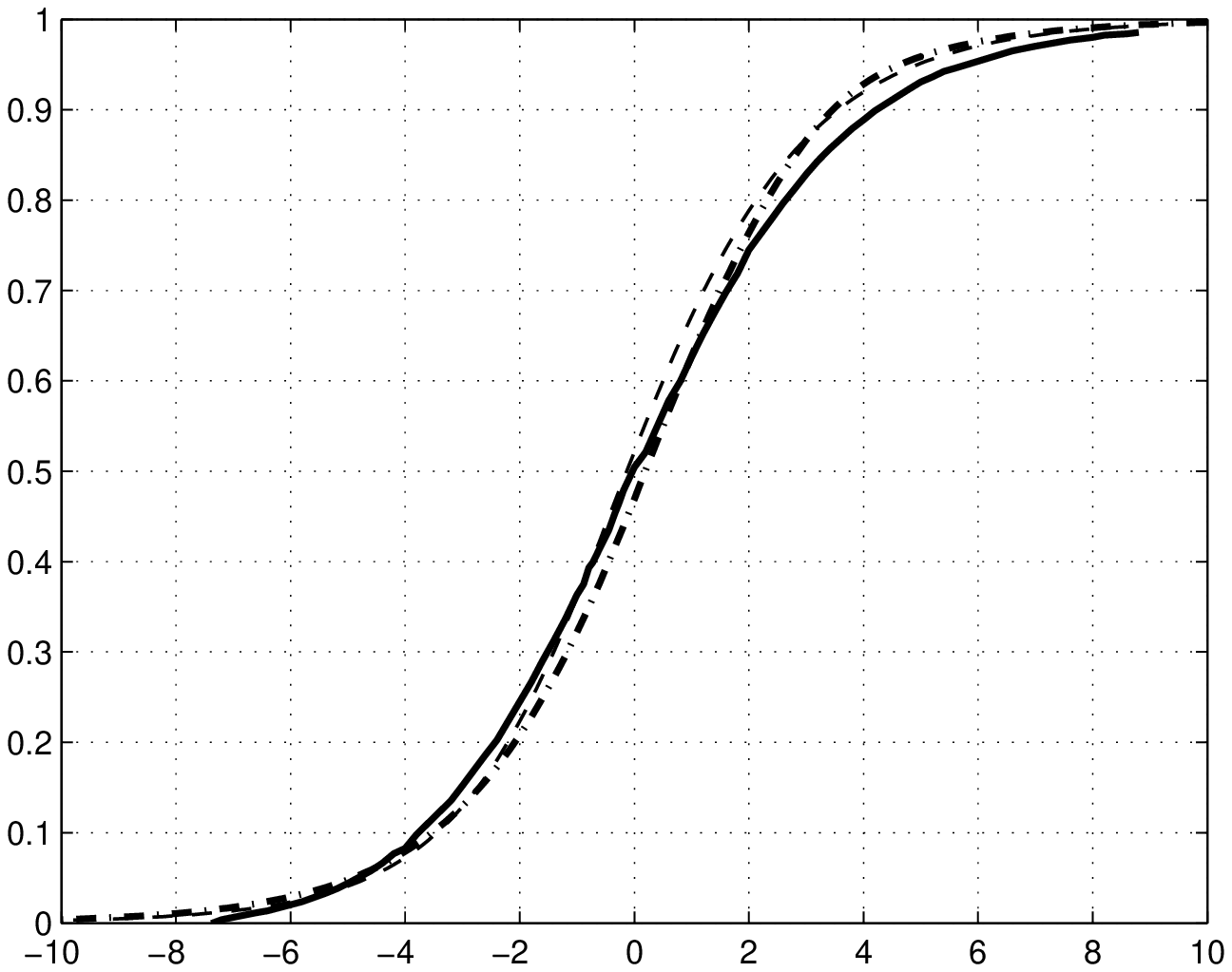}
}
\subfigure[$M_0 = 1.76$]{
   \includegraphics[scale=.45,clip]{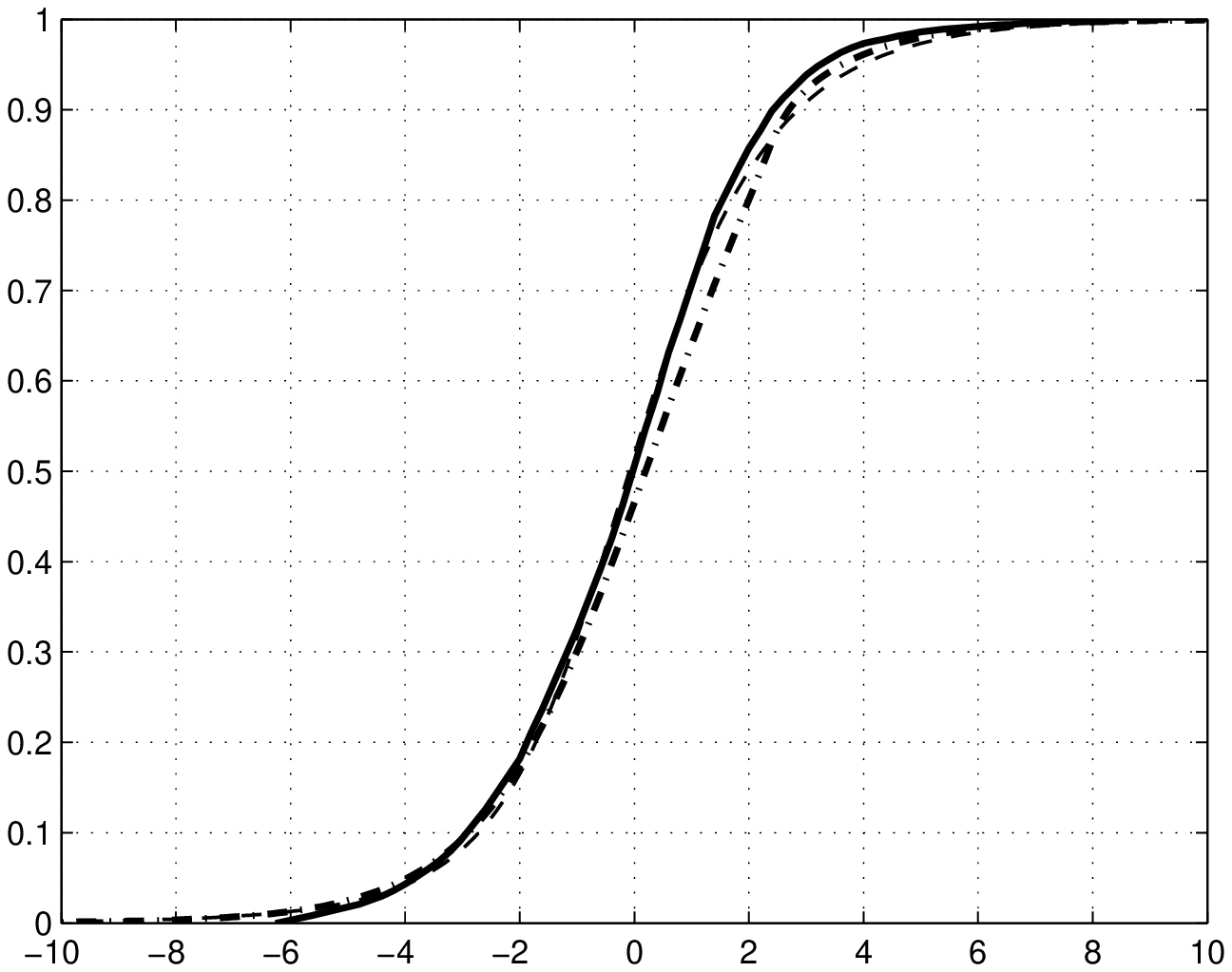}
}
\subfigure[$M_0 = 2.05$]{
   \includegraphics[scale=.45,clip]{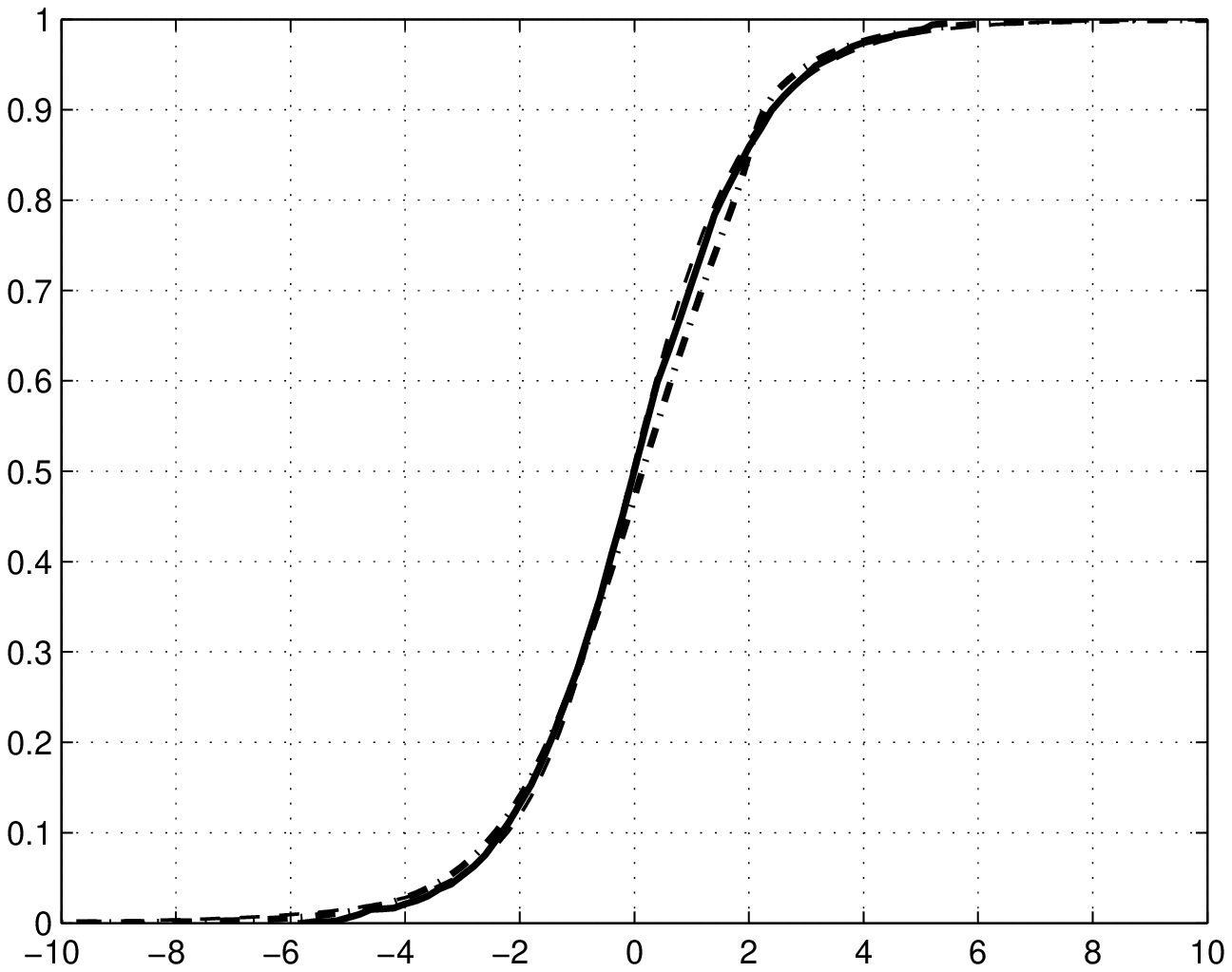}
}
\subfigure[$M_0 = 2.31$]{
   \includegraphics[scale=.45,clip]{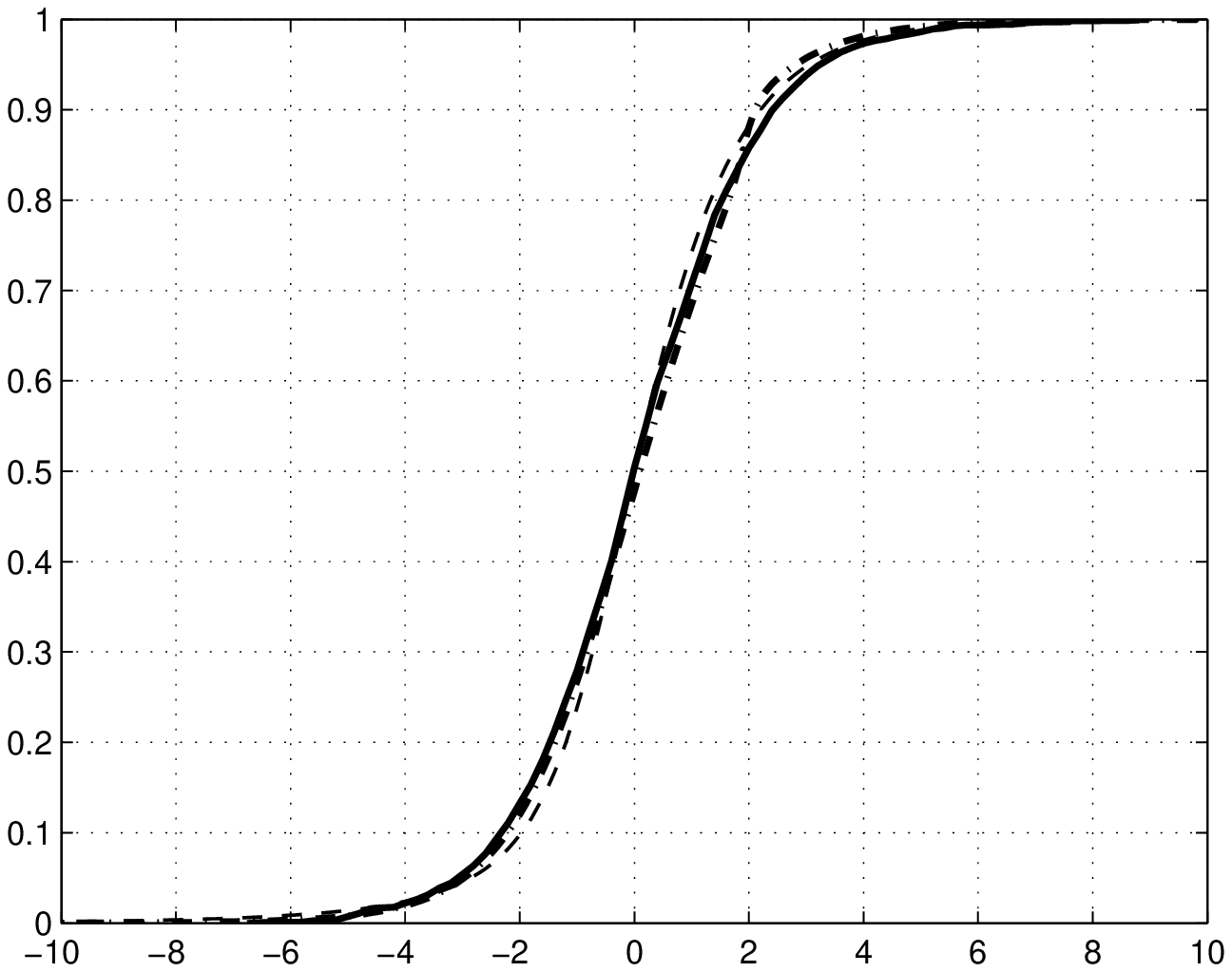}
}
\caption{The shock structure for different Mach numbers. All lines are
density profiles. The solid lines are the experimental data, and the
dashdot lines are R20 results. The thin dashed lines are the results
of the discrete velocity model (to be continued).}
\end{figure}

\addtocounter{figure}{-1}
\begin{figure}[!ht] 
\centering
\setcounter{subfigure}{4}
\subfigure[$M_0 = 3.38$]{
   \includegraphics[scale=.45,clip]{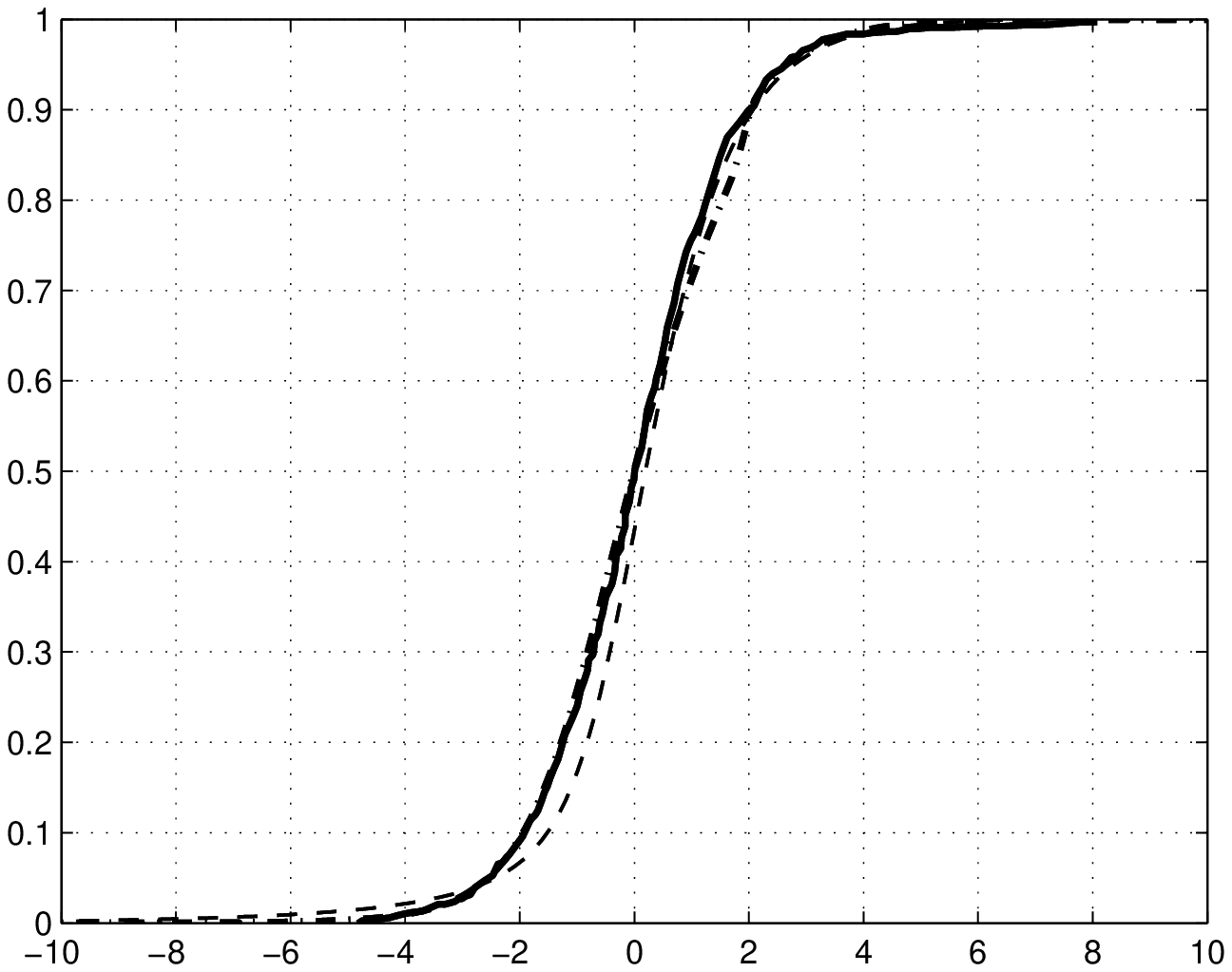}
}
\subfigure[$M_0 = 3.8$]{
   \includegraphics[scale=.45,clip]{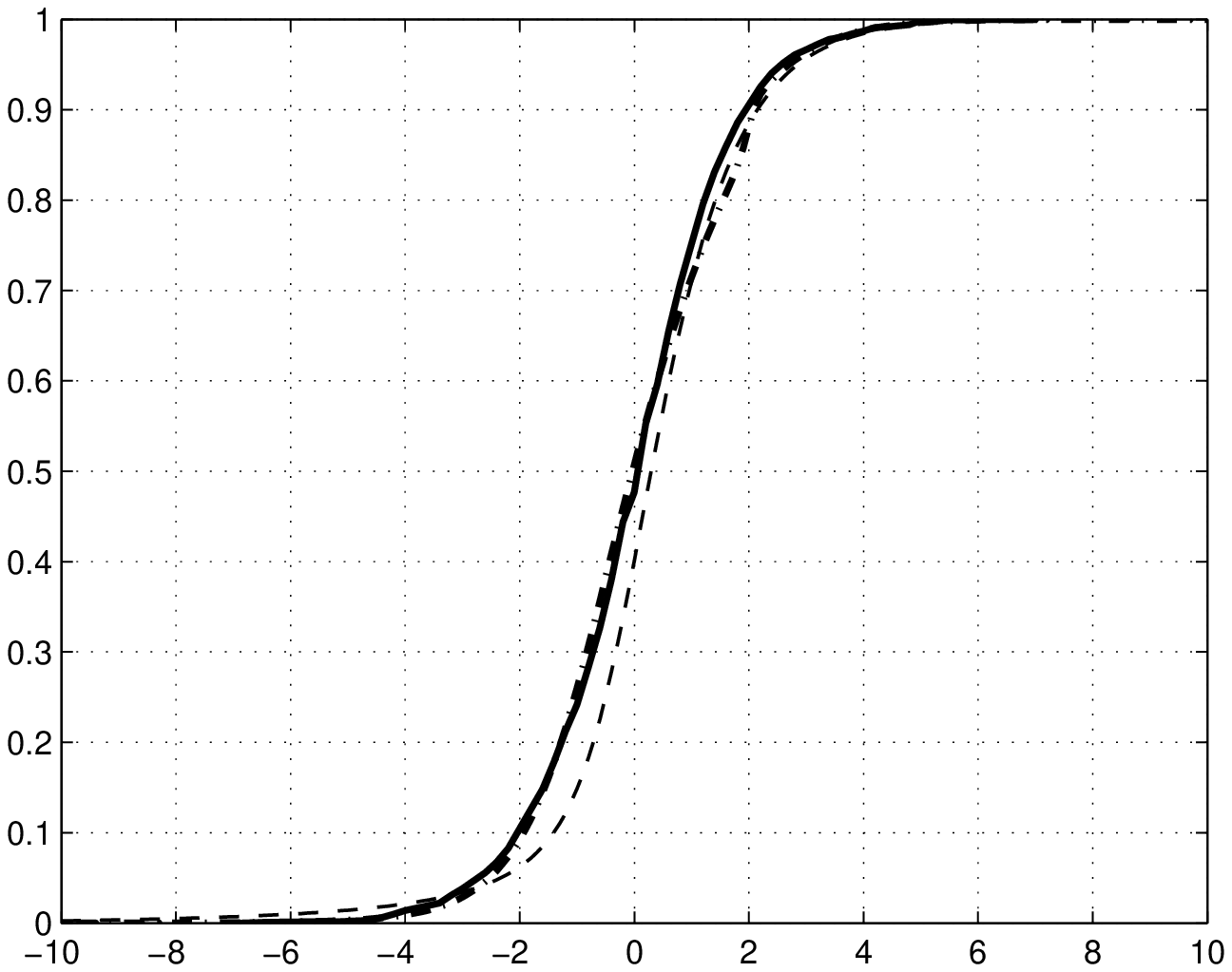}
}
\subfigure[$M_0 = 6.5$]{
   \includegraphics[scale=.45,clip]{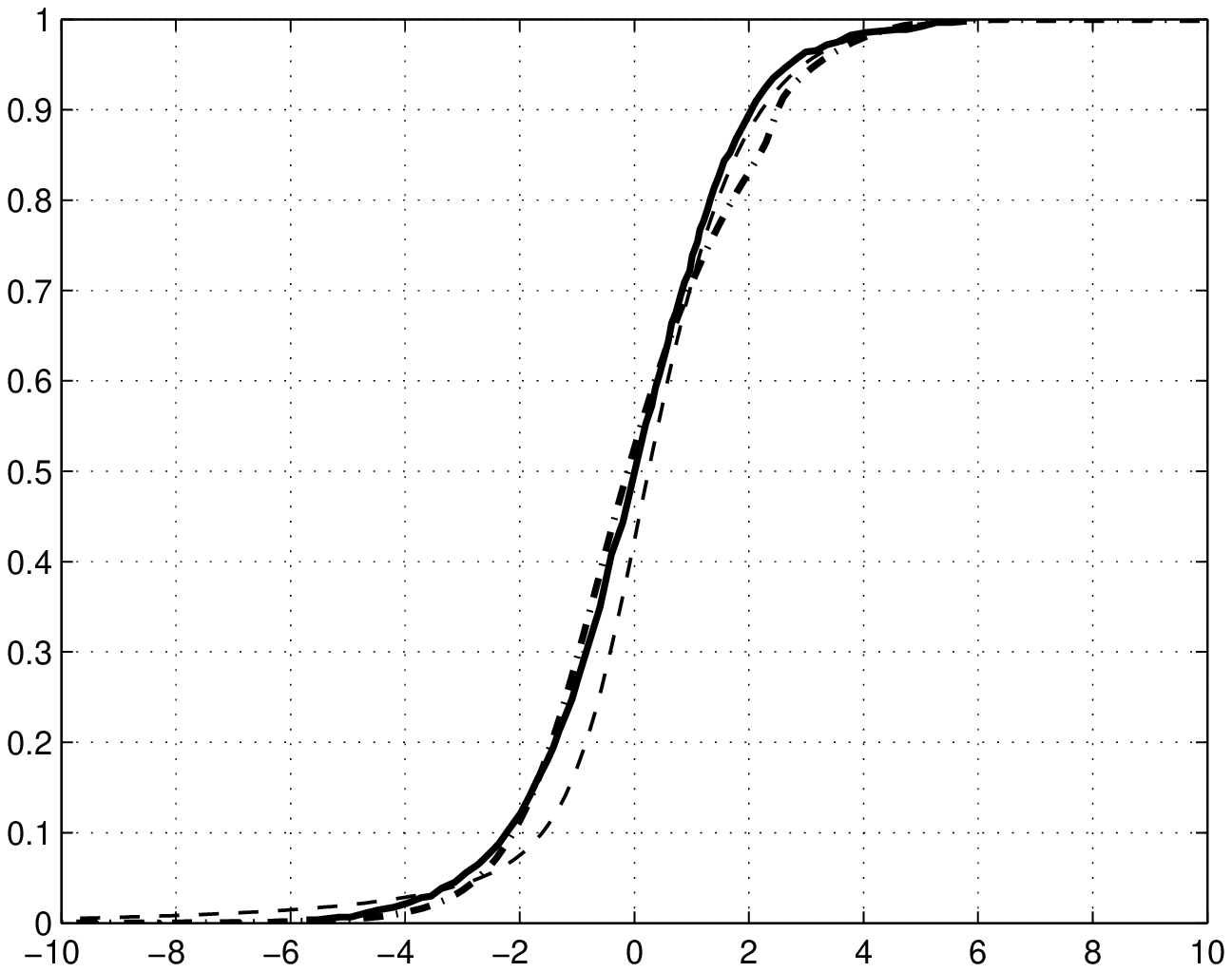}
}
\subfigure[$M_0 = 9.0$]{
   \includegraphics[scale=.45,clip]{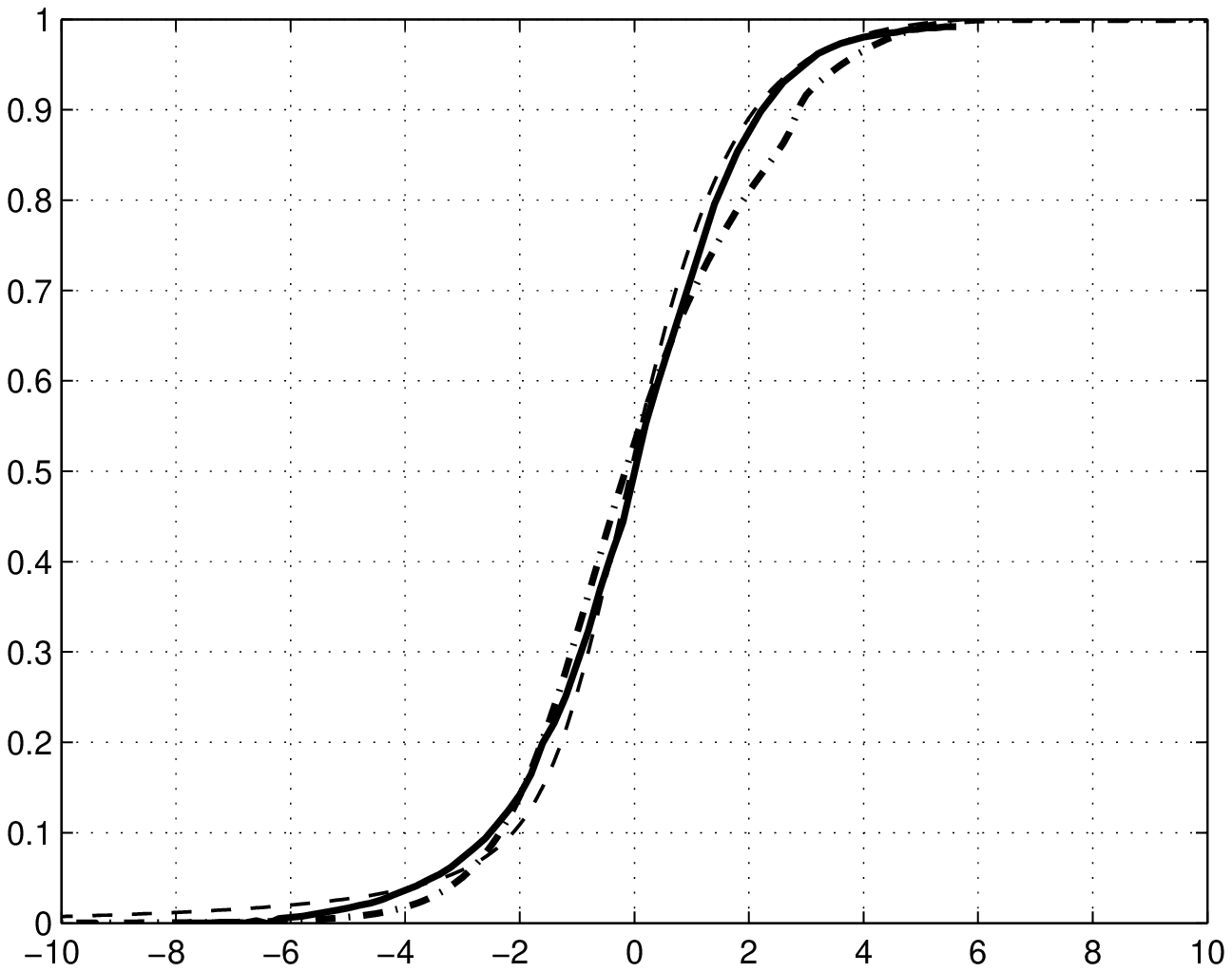}
}
\caption{The shock structure for different Mach numbers. All lines are
density profiles. The solid lines are the experimental data, and the
dashdot lines are R20 results. The thin dashed lines are the results
of the discrete velocity model.}
\label{fig:Shock-Structure}
\end{figure}

As stated in \cite{Muller}, in Grad's moment theory, a continuous
shock structure exists only up to the largest characteristic velocity.
For the case of $M = 3$, i.e. the 20-moment equations, Weiss
\cite{Shock} has found  that no continuous shock is possible when the Mach
number is larger than $1.808$. In Figure \ref{fig:Shock-Structure},
the moment system with the new regularization produces stable and
smooth shock structures for much greater Mach numbers. For $M_0 < 3$,
the R20 equations give good agreement with the experimental data,
while for larger Mach numbers, the profiles are generally correct,
although the predicted density is somewhat lower than the physical
case in the high density region.

%%% Local Variables: 
%%% mode: latex 
%%% TeX-master:"article" 
%%% End:

\section{Concluding remarks and discussions}
A numerical regularized moment method has been presented. In
order to construct the regularization term, we first use
Maxwellian iteration to determine the order of magnitude for
each moment, and then approximate the high order moments by
eliminating terms with small magnitude. Finally, the
approximation is greatly simplified by the strategy of
linearization. Compared with the regularization in
\cite{NRxx}, this method not only makes it possible to solve
high Mach number flow, but also keeps the convergence to the
BGK solution in moment number. Currently, it is still a
challenge to get physical shock profiles by this method, and
the work on a comprehensive analysis on the shock structure
of large moment systems is in progress.

\section*{Acknowledgements}
We thank the anonymous referee for his instructive remarks which
greatly improve the quality of the current paper. The research of
the second author was supported in part by the National Basic Research
Program of China (2011CB309704) and the National Science Foundation of
China under the grant 10771008 and grant 10731060.

%%% Local Variables: 
%%% mode: latex
%%% TeX-master: "article"
%%% End: 

% vim: tw=70:spell
\section*{Appendix}

\appendix

\section{Some properties of Hermite polynomials} \label{sec:Hermite}
The Hermite polynomials defined in \eqref{eq:He} are a set of
orthogonal polynomials over the domain $(-\infty, +\infty)$. Their
properties can be found in many mathematical handbooks such as \cite
{Abramowitz}. Some useful ones are listed below:
\begin{enumerate}
\item Orthogonality: $\displaystyle \int_{\bbR} \He_m(x)
\He_n(x) \exp(-x^2/2) \dd x = m! \sqrt{2\pi} \delta_{m,n}$;
\item Recursion relation: $\He_{n+1}(x) = x \He_n(x) - n \He_{n-1}(x)$;
\item Differential relation: $\He_n'(x) = n \He_{n-1}(x)$.
\end{enumerate}
And the following equality can be derived from the last two relations:
\begin{equation}
[\He_n(x) \exp(-x^2/2)]' = -\He_{n+1}(x) \exp(-x^2/2).
\end{equation}

\section{Derivation of the moment equations} \label{sec:moment_eqs}
In order to derive the analytical form of the moment equations, we
need to put the expanded distribution \eqref{eq:expansion} into the
Boltzmann-BGK equation \eqref{eq:BGK}. The subsequent calculation will
involve the temporal and spatial differentiation of the basis function
$\mathcal{H}_{\theta,\alpha}(\bv)$, which will be first calculated as
\begin{equation} \label{eq:dHds}
\begin{split}
  \pdd{s}\mathcal{H}_{\theta,\alpha}(\bv)& = -
    \frac{|\alpha|+D}{2}(2\pi)^{-\frac{D}{2}}
    \theta^{-\frac{|\alpha|+D}{2}-1}
    \pd{\theta}{s} \prod_{d=1}^D \He_{\alpha_d}(v_d) 
    \exp\left(-\frac{v_d^2}{2}\right) \\
  & \quad -(2\pi)^{-\frac{D}{2}}\theta^{-\frac{|\alpha|+D}{2}}
    \sum_{j=1}^{D} \left[\pd{v_{j}}{s} 
    \prod_{d=1}^{D} \He_{\alpha_{d} +\delta_{jd}}(v_d) 
    \exp\left(-\frac{v_d^2}{2}\right)\right] \\ 
  &= -\frac{|\alpha|+D}{2\theta} \pd{\theta}{s}
    \mathcal{H}_{\theta,\alpha}(\bv) -
    \sqrt{\theta}\sum_{d=1}^{D} \pd{v_{d}}{s}
    \mathcal{H}_{\theta,\alpha+e_{d}}(\bv),
\end{split}
\end{equation}
where $s$ stands for $t$ or $x_j$, $j = 1,2,3$. The partial derivative
$\partial v_d / \partial s$ can be expanded as
\begin{equation} \label{eq:dvds}
  \pd{v_d}{s} = \pdd{s}\left(\frac{\xi_d -u_d}{\sqrt{\theta}}\right)
    = -\frac{1}{\sqrt{\theta}}\pd{u_d}{s} -
      \frac{v_d}{2\theta} \pd{\theta}{s}.
\end{equation}
Noting that the recursion of the Hermite polynomials gives
\begin{equation} \label{eq:vH}
v_d \mathcal{H}_{\theta, \alpha+e_d}(\bv) =
  \sqrt{\theta} \mathcal{H}_{\theta, \alpha + 2e_d}(\bv) +
  \frac{\alpha_d + 1}{\sqrt{\theta}} \mathcal{H}_{\theta, \alpha}(\bv),
\end{equation}
we conclude from \eqref{eq:dHds}, \eqref{eq:dvds} and \eqref{eq:vH}
that
\begin{equation} \label{eq:dHds1}
\pdd{s} \mathcal{H}_{\theta,\alpha}(\bv) =
  \sum_{d=1}^D\pd{u_d}{s}
  \mathcal{H}_{\theta,\alpha+e_{d}}(\bv)+\frac{1}{2}\pd{\theta}{s}
  \sum_{d=1}^D \mathcal{H}_{\theta,\alpha+2e_d}(\bv).
\end{equation}
Replacing $s$ with $t$ in the above equation, one can have the
following expansion of the time derivative term in the Boltzmann-BGK
equation \eqref{eq:BGK} by some simple calculation:
\begin{equation} \label{eq:time_diff}
\begin{split}
\pd{f}{t} &= \sum_{\alpha \in \bbN^D} \left(
    \pd{f_{\alpha}}{t} \mathcal{H}_{\theta,\alpha} +
    f_{\alpha} \pd{\mathcal{H}_{\theta,\alpha}}{t}
  \right) \\
& = \sum_{\alpha \in \bbN^D} \left(
    \pd{f_{\alpha}}{t} + \sum_{d=1}^D \pd{u_d}{t} f_{\alpha-e_{d}}
    + \frac{1}{2} \pd{\theta}{t} \sum_{d =1}^D f_{\alpha-2e_d}
  \right) \mathcal{H}_{\theta,\alpha}.
\end{split}
\end{equation}
Now we consider the convection term. Substituting $x_j$ for $s$ in
\eqref{eq:dHds1}, and making use of \eqref{eq:vH} again, one has
\begin{equation} \label{eq:convection}
\begin{split}
\nabla_{\bx}\cdot(\bxi f) &= \sum_{j=1}^D \xi_j \pd{f}{x_j}
  = \sum_{j=1}^D (u_j + \sqrt{\theta} v_j)
    \sum_{\alpha \in \bbN^D} \left(
      \pd{f_{\alpha}}{x_j} \mathcal{H}_{\theta,\alpha} +
      f_{\alpha} \pd{\mathcal{H}_{\theta,\alpha}}{x_j}
    \right) \\
& = \sum_{\alpha \in \bbN^D}
  \mathcal{H}_{\theta,\alpha} \sum_{j=1}^D \bigg[ \left(
    \theta \pd{f_{\alpha-e_{j}}}{x_{j}} +
    u_{j}\pd{f_{\alpha}}{x_j} +
    (\alpha_{j}+1)\pd{f_{\alpha+e_j}}{x_{j}}
  \right) \\
& \qquad {} + \sum_{d=1}^D \pd{u_d}{x_j} \left(
  \theta f_{\alpha-e_d-e_j} + u_j f_{\alpha-e_d}
  + (\alpha_{j}+1) f_{\alpha-e_d+e_j}
\right) \\
& \qquad {} + \frac{1}{2} \pd{\theta}{x_j}
  \sum_{d=1}^D \left( \theta f_{\alpha-2e_d-e_j} +
  u_j f_{\alpha-2e_d} + (\alpha_{j}+1) f_{\alpha-2e_d+e_j}
\right) \bigg].
\end{split}
\end{equation}
Using $f_M = f_0 \mathcal{H}_{\theta,0}(\bv)$, the relaxation term can
be simply expanded as
\begin{equation} \label{eq:collision}
\frac{1}{\tau} (f_M - f) =
  -\frac{1}{\tau} \sum_{|\alpha| \geqslant 1}
  f_{\alpha} \mathcal{H}_{\theta,\alpha}(\bv).
\end{equation}

Finally, we combine \eqref{eq:time_diff}, \eqref{eq:convection} and
\eqref{eq:collision} and find the ultimate moment equations as
\begin{equation} \label{eq:moment_eqs}
\begin{split}
& \left( \pd{f_{\alpha}}{t} +
  \sum_{d=1}^D \pd{u_d}{t} f_{\alpha-e_{d}}
  + \frac{1}{2} \pd{\theta}{t} \sum_{d =1}^D f_{\alpha-2e_d}
  \right) \\
& \qquad + \sum_{j=1}^D \bigg[ \left(
  \theta \pd{f_{\alpha-e_{j}}}{x_{j}} +
  u_{j}\pd{f_{\alpha}}{x_j} +
  (\alpha_{j}+1)\pd{f_{\alpha+e_j}}{x_{j}}
\right) \\
& \qquad {} + \sum_{d=1}^D \pd{u_d}{x_j} \left(
  \theta f_{\alpha-e_d-e_j} + u_j f_{\alpha-e_d}
  + (\alpha_{j}+1) f_{\alpha-e_d+e_j}
\right) \\
& \qquad {} + \frac{1}{2} \pd{\theta}{x_j}
  \sum_{d=1}^D \left( \theta f_{\alpha-2e_d-e_j} +
  u_j f_{\alpha-2e_d} + (\alpha_{j}+1) f_{\alpha-2e_d+e_j}
\right) \bigg] = -\frac{1 - \delta_{0\alpha}}{\tau} f_{\alpha},
\end{split}
\end{equation}
where $\alpha \in \bbN^D$ and 
\begin{equation}
\delta_{0\alpha} = \left\{ \begin{array}{ll}
  1, & \alpha = 0, \\
  0, & \text{otherwise}.
\end{array} \right.
\end{equation}

\bibliographystyle{plain}
\bibliography{../article}

\begin{thebibliography}{10}

\bibitem{Abramowitz}
M.~Abramowitz and I.~A. Stegun.
\newblock {\em Handbook of Mathematical Functions with Formulas, Graphs, and
  Mathematical Tables}.
\newblock Dover, New York, 1964.

\bibitem{Alsmeyer}
H.~Alsmeyer.
\newblock Density profiles in argon and nitrogen shock waves measured by the
  absorption of an electron beam.
\newblock {\em J. Fluid. Mech.}, 74(3):497--513, 1976.

\bibitem{Bird}
G.~A. Bird.
\newblock {\em Molecular Gas Dynamics and the Direct Simulation of Gas Flows}.
\newblock Oxford: Clarendon Press, 1994.

\bibitem{NRxx}
Z.~Cai and R.~Li.
\newblock Numerical regularized moment method of arbitrary order for
  {B}oltzmann-{BGK} equation.
\newblock {\em SIAM J. Sci. Comput.}, 32(5):2875--2907, 2010.

\bibitem{Cai}
Z.~Cai, R.~Li, and Y.~Wang.
\newblock An efficient {\NRxx} method for {B}oltzmann-{BGK} equation.
\newblock {\em J. Sci. Comput.}, 2011.
\newblock DOI: 10.1007/s10915-011-9475-5.

\bibitem{Grad}
H.~Grad.
\newblock On the kinetic theory of rarefied gases.
\newblock {\em Comm. Pure Appl. Math.}, 2(4):331--407, 1949.

\bibitem{Holway}
L.~H. Holway.
\newblock New statistical models for kinetic theory: Methods of construction.
\newblock {\em Phys. Fluids}, 9(1):1658--1673, 1966.

\bibitem{Ikenberry}
E.~Ikenberry and C.~Truesdell.
\newblock On the pressures and the flux of energy in a gas according to
  {M}axwell's kinetic theory {I}.
\newblock {\em J. Rat. Mech. Anal.}, 5(1):1--54, 1956.

\bibitem{Mieussens}
L.~Mieussens.
\newblock Discrete velocity model and implicit scheme for the {BGK} equation of
  rarefied gas dynamics.
\newblock {\em Math. Models Methods Appl. Sci.}, 10(8):1121--1149, 2000.

\bibitem{Reitebuch}
I.~M{\"u}ller, D.~Reitebuch, and W.~Weiss.
\newblock Extended thermodynamics -- consistent in order of magnitude.
\newblock {\em Continuum Mech. Thermodyn.}, 15(2):113--146, 2002.

\bibitem{Muller}
I.~M{\"u}ller and T.~Ruggeri.
\newblock {\em Rational Extended Thermodynamics, Second Edition}, volume~37 of
  {\em Springer tracts in natural philosophy}.
\newblock Springer-Verlag, New York, 1998.

\bibitem{Shakhov}
E.~M. Shakhov.
\newblock Generalization of the {K}rook kinetic relaxation equation.
\newblock {\em Fluid Dyn.}, 3(5):95--96, 1968.

\bibitem{Struchtrup2004}
H.~Struchtrup.
\newblock Stable transport equations for rarefied gases at high orders in the
  {K}nudsen number.
\newblock {\em Phys. Fluids}, 16(11):3921--3934, 2004.

\bibitem{Struchtrup2005}
H.~Struchtrup.
\newblock Derivation of 13 moment equations for rarefied gas flow to second
  order accuracy for arbitrary interaction potentials.
\newblock {\em Multiscale Model. Simul.}, 3(1):221--243, 2005.

\bibitem{Struchtrup}
H.~Struchtrup.
\newblock {\em Macroscopic Transport Equations for Rarefied Gas Flows:
  Approximation Methods in Kinetic Theory}.
\newblock Springer, 2005.

\bibitem{Struchtrup2003}
H.~Struchtrup and M.~Torrilhon.
\newblock Regularization of {G}rad's 13 moment equations: Derivation and linear
  analysis.
\newblock {\em Phys. Fluids}, 15(9):2668--2680, 2003.

\bibitem{Torrilhon2006}
M.~Torrilhon.
\newblock Two dimensional bulk microflow simulations based on regularized
  {G}rad's 13-moment equations.
\newblock {\em SIAM Multiscale Model. Simul.}, 5(3):695--728, 2006.

\bibitem{Torrilhon_CiC}
M.~Torrilhon.
\newblock Hyperbolic moment equations in kinetic gas theory based on
  multi-variate {P}earson-{IV}-distributions.
\newblock {\em Commun. Comput. Phys.}, 7(4):639--673, 2010.

\bibitem{Shock}
W.~Weiss.
\newblock Continuous shock structure in extended thermodynamics.
\newblock {\em Phys. Rev. E}, 52(6):R5760--R5763, 1995.

\end{thebibliography}
\end{document}